\documentclass[article, showpacs,notitlepage,twocolumn]{revtex4-1} 
\usepackage{amsmath,amssymb,amsfonts, graphics,graphicx,color,multirow}

\newcommand{\beginsupplement}{%
        \setcounter{table}{0}
        \renewcommand{\thetable}{S\arabic{table}}%
        \setcounter{figure}{0}
        \renewcommand{\thefigure}{S\arabic{figure}}%
     }

\newcommand{\kappasp}{\kappa_{\mathrm{m}}}
\newcommand{\kappatw}{\kappa_{\mathrm{tw}}}
\newcommand{\tkappab}{\tilde{\kappa}_{\rm{b}}}
\newcommand{\txi}{\tilde{\xi}}
\newcommand{\bmm}{\mathbf{m}}
\newcommand{\bmd}{\mathbf{\hat{m}}}
\newcommand{\bt}{\mathbf{t}}
\newcommand{\bn}{\mathbf{n}}
\newcommand{\bN}{\mathbf{N}}

\newcommand{\rmd}{\,\mathrm{d}}
\newcommand{\p}{\partial}
\newcommand{\br}{\mathbf{r}}
\newcommand{\calF}{\mathcal{F}}
\newcommand{\calFm}{\mathcal{F}_{\rm{m}}}

\newcommand{\calFc}{\mathcal{F}_{\rm{int}}}
\newcommand{\tcalF}{\tilde{\mathcal{F}}}
\newcommand{\tcalFm}{\tilde{\mathcal{F}}_{\rm{m}}}

\newcommand{\calH}{\mathcal{H}}

\newcommand{\tgamma}{\tilde{\gamma}}
\newcommand{\kT}{\rm{k_{\rm{B}}T}}
\newcommand{\nm}{\rm{nm}}
\newcommand{\pN}{\rm{pN}}
\newcommand{\hatu}{d}
\newcommand{\kappab}{\kappa_{\rm{b}}}
\newcommand{\kappaa}{\kappa_{\rm{a}}}
\newcommand{\elleq}{\ell_{\rm{eq}}}

\begin{document}
\title{Role of the membrane for mechanosensing by tethered channels}
\author{Benedikt \surname{Sabass}}
\author{Howard A. \surname{Stone}} 
\affiliation{Department of Mechanical and Aerospace Engineering, Princeton University, Princeton, USA}
\pacs{87.16.dm, 87.14.ep, 87.15.kt}
\begin{abstract} 
Biologically important membrane channels are gated by force at attached tethers. Here, we generically characterize the non-trivial interplay of force, membrane tension, and channel deformations that can affect gating. A central finding is that minute conical channel deformation under force leads to significant energy release during opening. We also calculate channel-channel interactions and show that they can amplify force sensitivity of tethered channels.
\end{abstract}
\maketitle
\paragraph{Introduction.--}
The conversion of mechanical signals to a biochemical response is essential for living matter. One important class of mechanosensing proteins are membrane channels that are required for numerous biological functions, such as hearing~\cite{chalfie}, the sense of touch~\cite{martinac2004mechanosensitive}, or regulation of intracellular mechanics~\cite{hayakawa2008actin}. Recently, mechanosensitive channels have received considerable scientific attention, mostly with a focus on tension-sensing~\cite{bass2002crystal, sukharev2004mechanosensitive}. A fundamental insight was that membrane energy is sufficient to cause channel deformations that lead to gating, i.e. opening or closing of the channel. As reviewed in~\cite{phillips2009emerging}, the deformation modes include conical shape changes~\cite{dan1998effect, turner2004gating,lee2006energetics, chen2008gating, rautu2015membrane}, radial expansion~\cite{wiggins2004analytic, markin2004thermodynamics,reeves2008membrane, pak2015gating}, or changes in channel hydrophobic thickness~\cite{huang1986deformation, helfrich1990calculation}. However, in many cases the channels are directly tethered to cytoskeletal or extracellular structures, which allow a direct transmission of mechanical force~\cite{kung2005possible, orr2006mechanisms, martinac2014ion}. 

One example of a tethered channel is the DEG/ENaC complex, which conveys touch sensing in C. elegans~\cite{tavernarakis1997molecular}. Here, an ion channel is likely opened by mechanical interaction with intracellular or extracellular proteins~\cite{o2005mec}. Further examples of tethered channels are force-sensitive TRP channels that possess intracellular ankyrin domains. Ankyrin repeats are proposed to function as ``gating springs'' that convey force~\cite{howard2004hypothesis,delmas2013mechano, liu2015forcing, zhang2015ankyrin}. These tethers have an estimated stiffness of $1\,\pN/\nm$ and a working range on the order of $10\,\nm$. Thus, forces are estimated to be around $10\,\pN$~\cite{sotomayor2005search, li2006stepwise}. The $\pN$ force scale is also confirmed by experiments~\cite{hayakawa2008actin}. Since the observed gating is stochastic~\cite{prager2014unique, zhang2015ankyrin}, energy barriers are expected to be comparable to the thermal energy. 

A role of the membrane has been experimentally verified for tethered TRPA1 channels. Here, gating depends robustly and asymmetrically 
on membrane curvature that is induced by amphipathic molecules, partitioning either in the inner or outer leaflet~\cite{hill2007trpa1}. Furthermore, GsMTx-4, a toxin that inhibits tension-activated channels through perturbing the bilayer~\cite{suchyna2004bilayer}, causes gating of TRPA1~\cite{hill2007trpa1}. In spite of experimental evidence, a theoretical analysis of the role of the membrane for tethered channels is lacking, and this is the focus of this Letter.

In quasi-equilibrium, channel gating is governed by an energy $\calF$, which depends on the internal molecular state and on the deformation of the membrane around the channel. Thus, $\calF = \calFc + \calFm$. The internal energy $\calFc$ is determined by structural details, whose characterization requires intricate molecular dynamics studies~\cite{khalili2009molecular}. In contrast, the membrane energy $\calFm$ always affects gating if sufficient channel shape changes occur~\cite{phillips2009emerging}. Focusing on generic principles, we study how force and the membrane affect two main channel deformation modes, namely conical deformation and radial expansion. 
\begin{figure}[htb]
  \centering
	\includegraphics[scale=0.91]{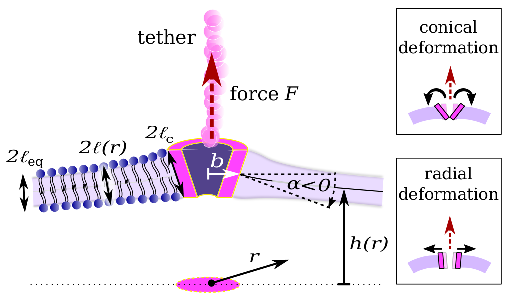} 
\caption{Model for a tethered channel embedded in a lipid bilayer. Variables are defined 
in the main text. The channel shape changes either by conical deformation where the angle 
$\alpha$ varies or by radial deformation with variation of $b$.} 
\label{fig_channel_model}
\end{figure}
\paragraph{Calculation of membrane energy.--}We consider a radially symmetric channel that is placed on the centerline of a cylindrical system with radial coordinate $r$ (Fig.~\ref{fig_channel_model}). A constant vertical force $F$ is exerted on the channel, with signs chosen such that $F>0$ is directed upwards.
The channel radius is denoted by $b$; typically $b \simeq 3\,\nm$. A conical channel shape is characterized by the angle $\alpha$, with signs chosen such that $\alpha >0$ corresponds to a channel with small side pointing upwards. The channel is surrounded by a fluid lipid bilayer. The height of the center of the bilayer above a reference plane is denoted by $h(r)$, with $h(0)$ being the height at the channel center. The bilayer thickness is denoted by $2\ell(r)$. Hydrophobic properties of the channel can force the membrane leaflets to splay, leading to a perturbation of the equilibrium leaflet thickness $u(r)\equiv \ell-\ell_{\rm{eq}}$. Typically, $\ell_{\rm{eq}} \simeq 1.75\,\nm$~\cite{rawicz2000effect}. The linearized membrane deformation energy is~\cite{huang1986deformation,Suppl_mat}
\begin{align}
\begin{split}
&\calFm = \int{\left[\frac{\kappa_{\rm{b}}}{2}(\nabla^2 u)^2 + \frac{\kappa_{\rm{a}}}{2} \left(\frac{u}{\ell_{\rm{eq}}}\right)^2+
\frac{\gamma}{2}(\nabla u)^2\right]\rmd^2r}\\
&+\int{\left[\frac{\kappa_{\rm{b}}}{2}(\nabla^2 h)^2 +\frac{\gamma}{2}(\nabla h)^2\right]\rmd^2r}+\gamma\int{\rmd^2r} - F h(0),\label{eq_G_full}
\end{split}
\end{align}
where the surface integral extends over the entire reference plane outside the channel. Neglecting shear forces between the leaflets, we assume that the bending modulus $\kappab$ is the same for thickness perturbations $u$ and height perturbations $h$. Typically, $\kappab \simeq 25\,\kT$~\cite{rawicz2000effect}. Changes of membrane thickness are penalized by the term $\sim  \kappaa u^2$, where $\kappaa \simeq 40\,\kT/\nm^2$~\cite{rawicz2000effect,huang1986deformation}. Tension $\gamma$ maintains constant area of both leaflets. For eukaryotes, tension is usually low, $\gamma \simeq 10^{-3}\,\kT/\nm^2$~\cite{dai1999membrane}, and large amounts of excess area are believed to lead to constant tension~\cite{raucher1999characteristics}. Therefore, in line with previous research~\cite{derenyi2002formation}, we assume that the force $F$ does not appreciably affect membrane tension. A large natural scale for tension can be fixed by combining the bending modulus and channel radius as $\gamma_S\equiv\kappab/b^2$; typically $\gamma_S\simeq2.7\,\kT/\nm^2\gg \gamma$.

Depending on membrane composition, the orientational ordering of lipids may affect deformations on the nanometer scale~\cite{jablin2014experimental, jensen2004lipids, andersen2007bilayer}, which may require further terms in	Eq.~(\ref{eq_G_full})~\cite{mackintosh1991orientational,seifert1996role,fournier1999microscopic, kozlovsky2004orientation,may2004tilt,kuzmin2005line, venturoli2006mesoscopic,watson2013thermal, argudo2016continuum}. The influence of lipid tilt is analyzed in the Supplemental Material. Furthermore, elastic interactions between the membrane and environment may change $\calFm$. An elastically supported membrane is studied in the Supplemental Material.

The membrane height $h(r)$ is locally determined by a combined effect of tension and bending, which allows the introduction of a characteristic lengthscale as
\begin{align}
\xi^{-1}\equiv\sqrt{\kappa_{\rm{b}}/\gamma}. \label{eq_def_xi}
\end{align}
Typical values of the lengthscale are $\xi^{-1}\simeq [5 \ldots 500]\,\nm$, which is larger than the channel radius $b$. 
A variation of Eq.~(\ref{eq_G_full}) yields equilibrium equations determining $h(r)$ as
\begin{subequations}
\begin{align}
\nabla^2(\nabla^2 -\xi^2)h = 0, & &\\
F= 2 \pi b \kappab \p_{r}\left(\nabla^2 h -\xi^2 h\right)|_{b}, & & \partial_r h|_{b} =\alpha,
\end{align}
\end{subequations}
where the boundary conditions involving $F$ and $\alpha$ fix the height and contact angle at the channel. Far away
from the channel, at $r=L\gg b$, we assume $h|_{L} =0$. 

A variation of Eq.~(\ref{eq_G_full}) also yields the equation determining thickness perturbations $u(r)$.
If the membrane leaflets are fixed to the channel walls by chemical interactions, the force $F$ does not change
the thickness perturbation around the channel. Then we have 
\begin{subequations}
\begin{align}
(\nabla^2 -\eta^2_{-})(\nabla^2 -\eta^2_{+})u=0, & & \\
u|_{b} =(\ell_{c} - \ell_{\rm{eq}}),& & \partial_r u|_{b} =0,
\end{align}
\end{subequations}
where $\eta^2_{\pm} \equiv \left(\xi^2 \pm \sqrt{\xi^4 - 4\kappaa/(\elleq^2 \kappab)}\right)/2$
and $\partial_r u|_{L} =0$, $u|_{L} =0$. The lengthscale of thickness perturbations is $\bar{\eta}^{-1} \equiv (\eta_{+}^{-1}+\eta_{-}^{-1})$, which is about $1.5\,\nm$~\cite{nielsen1998energetics, ursell2008role}. 

Once $u$ and $h$ are calculated, the energy $\calFm$ results from Eq.~(\ref{eq_G_full})~\cite{Suppl_mat}. Due to its two-dimensional nature, $\calFm$ displays a logarithmic divergence with system size $L$. However, this divergent energy does not depend on channel shape parameters, and is thus immaterial for gating. We remove the divergence and other constants by defining $\tcalFm \equiv \calFm + F^2\,[\Gamma_e +\log(\xi L/2)]/(4 \pi \gamma)$, where $\Gamma_e=0.577\ldots$ is the Euler-Mascheroni constant. Since usually $b\xi < 1$, the membrane energy can be expanded up to $O((b\xi)^{3})$ to yield a transparent formula
\begin{align}
\tcalFm \approx \hatu^2 b-\gamma \pi b^2(1- \alpha^2 X) +b \alpha F X  + \frac{b^2 (F + 2 b \alpha \gamma \pi)^2 Y}{4 \kappa_{\rm{b}} \pi},\label{eq_Overall_energy_simpl}
\end{align}
where $\hatu^2 \equiv \pi \kappa_{\rm{b}}(\ell_{c}-\ell_{\rm{eq}})^2 \eta_{+}\eta_{-}(\eta_{+}+\eta_{-})$~\cite{Suppl_mat}. Further constants are $X \equiv -\Gamma_e - \log[b \xi/2]$ and $Y\equiv (1 + 2 X + 2 X^2)/4$; $X$ and $Y$ are both positive for $b\xi<1$. 

Eq.~(\ref{eq_Overall_energy_simpl}) allows to calculate how membrane energy changes with channel shape. The first term $\hatu^2 b$ results from membrane thickness perturbations $u$ around the channel. This energy is independent of $F$ and penalizes radial expansion. Note that $\hatu^2$ may change with channel deformation if the hydrophobic thickness varies. Other small-scale effects, such as tilt of the lipids, bending of their acyl chains, and the detailed channel structure may also affect $\hatu^2$. The second term in Eq.~(\ref{eq_Overall_energy_simpl}) results from membrane tension. This energy contribution usually decreases with radius $b$, except when the channel is very conical $\alpha^2 X > 1$. The two last terms in Eq.~(\ref{eq_Overall_energy_simpl}) contain the effect of the force. An important role of the conical shape is evident from the occurrence of $\alpha$ in both terms. In the following, we use $\calFm$ to analyze the energetics of channel deformation.
\begin{figure}[htb]
  \centering
	\includegraphics[scale=0.87]{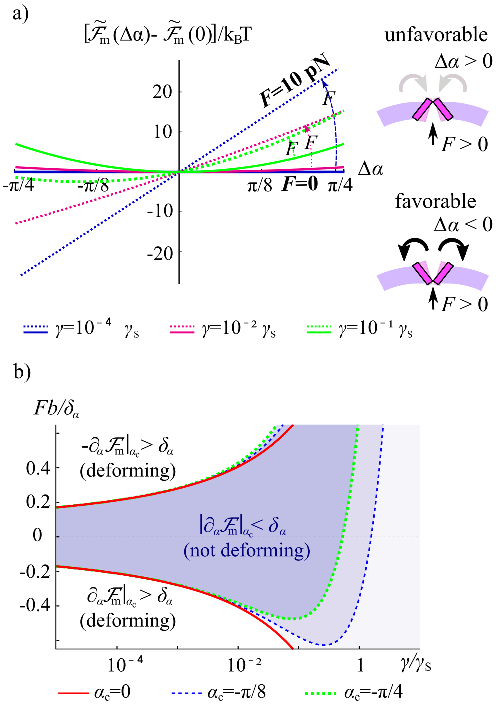} 
\caption{a)~Change of membrane energy with conical angle $\Delta\alpha$ when the channel is initially cylindrical $(\alpha_c=0)$. Force $F>0$ tilts the function to make deformations $\Delta\alpha<0$ energetically favorable (dotted lines). b)~Lines of threshold forces $F^{*}(\gamma, \alpha_c)$ that allow opening against an internal molecular resistance $\delta_{\alpha}/b$. Note the amplification of applied force $F^{*} b/\delta_{\alpha} < 1$, which results from membrane leverage. 
Representative parameters: $b=3\,\nm$, $\kappab = 25\,\kT$, $\delta_{\alpha} = 180\,\kT/\pi$.} 
\label{fig_angular_def}
\end{figure}
\paragraph{Conical deformation.--}We first study a change of the conical angle $\alpha$. For an initially closed channel with $\alpha =\alpha_c$, the angle may change during gating as $\alpha_c\rightarrow\alpha_c+\Delta\alpha$. The membrane favors such a channel deformation if the energy is reduced as $\calFm(\alpha_c +\Delta \alpha)-\calFm(\alpha_c) <0$. Fig.~\ref{fig_angular_def}a indicates how $\calFm$ changes with $\Delta\alpha$. A force $F$ tilts the energy function, making conical deformation favorable in one angular direction and unfavorable in the other. For physiologically low membrane tension $<10^{-1} \gamma_S$, Fig.~\ref{fig_angular_def}a illustrates that membrane energy varies almost linearly with $\Delta\alpha$. In this case, Eq.~(\ref{eq_Overall_energy_simpl}) yields
\begin{align}
\tcalFm(\Delta\alpha)-\tcalFm(0) \simeq b F X \,\Delta\alpha.\label{eq_turut}
\end{align}
Using this simple formula with typical values $F=10\,\pN$, $b=3\,\nm$, $\xi^{-1}=50\,\nm$, we find that an angular deformation of $\Delta\alpha = 3^{\circ}$ corresponds to an energy change of $\simeq1\,\kT$. Hence, minute molecule deformations of even $1\,\mathring{A}$ significantly affect the membrane energy in the presence of a force. 

Next, we assume that the molecular channel structure poses an energetic barrier to conical deformations. Without knowledge of details, the resistance to deformation can be described through an energy scale $\delta_{\alpha} \equiv \p_{\alpha}\calFc|_{\alpha_c}$. Applying force to a channel causes conical deformations if the net energy is reduced $-|\p_{\alpha}\calFm|_{\alpha_c} + \delta_{\alpha}\leq0$. Force thresholds $F^{*}$ for the occurrence of conical deformation are thus calculated from the criterion $|\p_{\alpha}\calF|_{\alpha_c}=\delta_{\alpha}$. Fig.~\ref{fig_angular_def}b displays lines for $F^{*}$ that separate parameter regions where conical deformation occurs. Note the scale on the ordinate $F b/\delta_{\alpha} < 1$, which means that a small applied force $F$ can overcome larger resisting internal force $\delta_{\alpha}/b$ and thus cause conical deformation. 

To understand this amplification of the force $F$ we use Eq.~(\ref{eq_turut}) and estimate $|\p_{\alpha}\calFm|_{\alpha_c} \sim b F X$. 
Equating membrane deformation energy with internal energy $|\p_{\alpha}\calFm|_{\alpha_c} =\delta_{\alpha}$, the threshold force follows 
as $F^{*} \sim \delta_{\alpha}/(b X)$. Thus, forces result from dividing the internal energy scale $\delta_{\alpha}$ by a lever arm length $b X\sim -b\log(b \xi)$, which includes the large scale membrane deformation. Since $X$ depends only logarithmically on $\gamma$ and $\kappab$, threshold forces are relatively robust against variation of these membrane properties. 

In Fig.~\ref{fig_angular_def}b, data for $\alpha_c\neq 0$ illustrates deformation of a channel that already has a conical shape. 
For small tension, $\gamma/\gamma_S \lesssim 10^{-2}$, all curves lie on top each other since force thresholds $F^{*}$ are not affected by the initial conical angle $\alpha_c$. On the other hand, strong tension can deform a channel with $\alpha_c\neq 0$ even when $F=0$.
\begin{figure}[htb]
  \centering
	\includegraphics[scale=0.87]{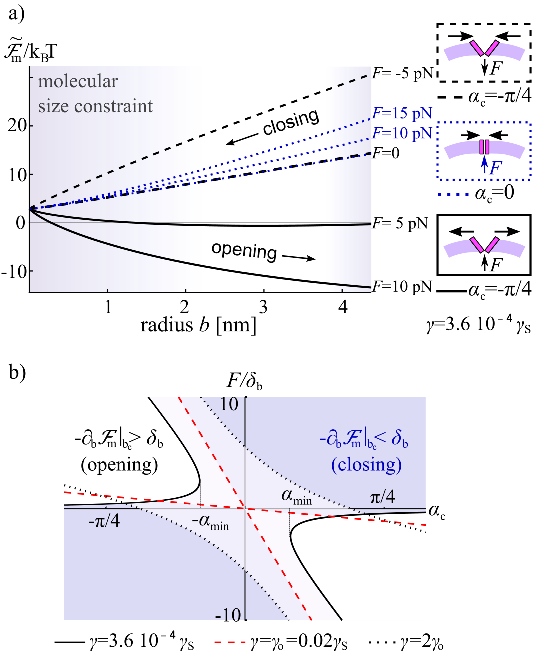} 
\caption{a)~Membrane energy $\tcalFm$ vs. channel radius $b$. Opening is favored when $\p_b\tcalFm<0$.
For $F=0$, an assumed hydrophobic mismatch $(\ell_{c} -\ell_{\rm{eq}})=0.2\,\nm$ dominates $\calFm$, thus $\p_b\tcalFm>0$. $\p_b\tcalFm<0$ requires a conical channel with $\alpha_c F < 0$. b)~Lines of threshold forces $F^{**}(\gamma, \alpha_c)$ at which radial expansion 
becomes possible. Full lines: For $\gamma<\gamma_o$, threshold forces describe hyperbolic regions where opening occurs when $|\alpha_c|>|\alpha_{\mathrm{min}}|$. Dashed and dotted lines: When $\gamma\geq\gamma_o$, membrane tension is sufficient to radially open a channel at $\alpha_c=0$. In b) $(\ell_{c} -\ell_{\rm{eq}})=0$. Representative parameters:~$\kappab= 25\,\kT$, $\kappaa = 40\,\kT/\nm^2$, $b_c= 3\,\nm$, $\delta_b = 1\,\kT/\nm$.} 
\label{fig_radial_def}
\end{figure}
\paragraph{Radial deformation.--}
Channel gating may lead to an increase of the radius $b$. However, if $\p_b\tcalFm(b) >0$, membrane deformation does not favor this expansion. Fig.~\ref{fig_radial_def}a shows the dependence of membrane energy $\tcalFm$ on radius $b$ for typical parameter values. Clearly, $\p_b\tcalFm(b) <0$ only occurs for a pronounced conical shape. For small angles $\alpha_c \simeq 0$ or large $F$, force always favor radial closure of the channel.

Analogous to the analysis of conical deformations, we next assume that the channel itself resists radial deformation through an internal force $\delta_{b} \equiv \p_b\calFc|_{b_c}$, which is caused by conformational changes. Threshold forces for radial opening $F^{**}$ are calculated from the condition $\p_b\calFm|_{b_c} + \delta_b = 0$. Fig.~\ref{fig_radial_def}b displays thresholds $F^{**}$ that separate regions of parameters $(F,\alpha_c)$ where channels are open or closed. The shape of these regions depends on tension $\gamma$. The critical tension $\gamma_o \equiv (\hatu^2+\delta_b)/(2 \pi b_c)$, which opens a cylindrical channel $\alpha_c=0$ when $F=0$, allows discrimination of two regimes. For $\gamma >\gamma_o$, we find one central parameter region where the channel is held open by tension. When $\gamma<\gamma_o$, two hyperbolic regions exist where force can lead to radial opening when pushing towards the larger side of a conical channel. These hyperbolic regions are limited by finite angles $|\alpha_{\mathrm{min}}|$. To $O((b\xi)^2)$, we find $\alpha_{\mathrm{min}}^2 \approx |\hatu^2 +\delta_b| b X^2 / (\kappab \pi (1 - X)^2)$. Since $\alpha$ is always limited by geometry, we can use this formula to estimate maximum internal forces $\delta_b$ that can be overcome by external forcing. Assuming $|\alpha_{\mathrm{min}}| <\pi/3$, $\hatu^2=0$ and typical parameters employed above, we estimate $\delta_b \lesssim 16\,\kT/\nm$. This maximum force scale is not large and membrane channels possibly have a less pronounced conical shape. Therefore, radial expansion in a weak-tension membrane is likely not favored by force, which is in contrast to conical deformation.
\paragraph{Interaction between channels.--}Membrane deformation can lead to collective effects where force at one channel affects the gating of neighboring channels. For $F=0$, the interaction between membrane inclusions has been studied extensively~\cite{goulian1993long, netz1995inhomogeneous, weikl1998interaction, fournier1999microscopic, chou2001statistical, evans2003interactions, muller2005geometry, haselwandter2013directional, fournier2014dynamics}. We consider here two channels with radii $b_{\{1,2\}}$ and conical angles $\alpha_{\{1,2\}}$ that are separated by a distance $R$ in a homogeneous membrane. A force $F$ is applied to each channel. The membrane energy $\tcalFm$ can be calculated approximately through a multipole expansion assuming $\xi b_{\{1,2\}} \ll 1$ and $R \gg b_{\{1,2\}}$, which is appropriate for $R \gtrsim 3 b_1$ when $b_1=b_2$~\cite{reynwar2011membrane}. For the individual channels, we write the deformation energy given by Eq.~(\ref{eq_Overall_energy_simpl}) as $\tcalF_{\rm{m},1}$ and $\tcalF_{\rm{m},2}$, where the parameters $b$, $\alpha$ are substituted by $b_{\{1,2\}}$, $\alpha_{\{1,2\}}$. For the interaction energy $\tcalF_{\rm{m}, R} \equiv \tcalFm - \tcalF_{\rm{m},1}-\tcalF_{\rm{m},2}$ we find~\cite{Suppl_mat} 
\begin{align}
\begin{split}
&\tcalF_{\rm{m}, R}\approx(\alpha_{1} b_{1} + \alpha_{2} b_{2}) F K_0(\xi R)+2 \pi \gamma b_{1} b_{2} \alpha_{1} \alpha_{2} K_0(\xi R)\\
&+\sum_{i=1,2} F^2 b_{i}^2\left(\frac{(1 + 2 X_{i})\,K_0(\xi R) }{16 \kappab \pi} -\frac{(1 - R \xi K_1(\xi R))^2}{8 \pi R^2 \gamma }\right),\label{eq_interaction_approx}
\end{split}
\end{align}
where $X_i = -\Gamma_e -\log(b_i \xi/2)$ and contributions of $O(b_i^2 \xi/R, (b_i \xi)^2)$ as well as terms that do not depend on $\alpha_i, b_i$ are neglected.  $K_n(x)$ are modified Bessel functions of the second kind. Interaction becomes significant when the distance $R$ is smaller than the lengthscale $\xi^{-1}$.

To study the role of interactions for gating we focus on the simplest case, namely conical deformation of initially cylindrical channels ($\alpha_c = 0$ for both channels). Since both forces have the same sign, interaction increases the energy that can be released by conical deformation. 
Analogous to the analysis for a single channel, force thresholds $F^{*}_2$ for combined conical deformation of two neighboring channels can be calculated from the condition $-|\p_{\alpha_i}\calFm|_{0}+\delta_{\alpha} =0$, $i\in\{1,2\}$. 
\begin{figure}[htb]
  \centering
	\includegraphics[scale=0.97]{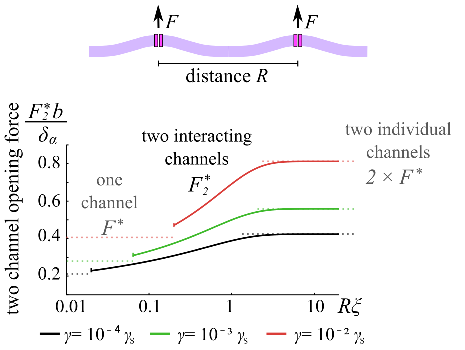} 
\caption{Channel interaction can enhance force sensitivity. For close distances, $R\xi \ll 1$, the force necessary for conical deformation of two channels $F^{*}_2$ is almost as small as for a single channel $F^{*}$. Lines end at the minimum distance $R=(b_1+b_2)$. $b_{\{1,2\}} =3\,\nm$, $\alpha_c=0$, $\kappab=25\,\kT$, $\delta_{\alpha} = 180\,\kT/\pi$.} 
\label{fig_cooperativity}
\end{figure}
Fig.~\ref{fig_cooperativity} demonstrates that force thresholds for conical deformation are significantly reduced by channel interaction. When $R \ll \xi^{-1}$, the force $F^{*}$ that is necessary to deform a single channel almost suffices to deform two channels. Although membrane-mediated interactions are not pairwise additive, two-channel interaction is believed to be dominant for sparsely distributed proteins~\cite{chou2001statistical}. Consequently, we expect from Eq.~(\ref{eq_interaction_approx}) that force-sensitivity of a channel ensemble can be amplified by collective mechanics, which would allow a response to weak, local forces.
\paragraph{Experiments and predictions.--}
The theoretical framework laid out in this letter is generic since the membrane energy~(\ref{eq_G_full}) affects 
gating of any deforming channel. Whether membrane energy dominates the gating process must be investigated for specific channels 
through experiments or molecular dynamics simulations. For both avenues, the theory provides helpful predictions and tools.

The current experimental status allows a few consistency tests. First, molecular structures~\cite{paulsen2015structure, cao2013trpv1} indicate that radial deformation of known tethered channels is small, order $0.1\,\nm$, but transmembrane units tilt during opening. This finding is in line with the above analysis, where conical channel deformations under force are favored by the membrane. Second, the observed activation of tethered TRP-channels through bilayer perturbations~\cite{hill2007trpa1} supports a role of the membrane.
Quantitative measurements of this type can be analyzed using Eq.~(\ref{eq_Overall_energy_simpl}) by calculating gating probabilities, which we describe in~\cite{Suppl_mat}. Third, the theory predicts that channel-channel interaction can lead to a sub-linear
dependence of opening force on the number of tethered channels. This effect may be measurable at small tensions, e.g., by force application with an optical trap~\cite{hayakawa2008actin} when the tether density is varied biochemically~\cite{prager2014unique}.

Finally, we emphasize that tension sensing and force sensing via tethers are not mutually exclusive, but complimentary mechanisms. While tension sensing requires $\gamma \gtrsim 1\,\kT/nm^2$, force sensing requires $\gamma \lesssim 0.1\,\kT/nm^2$ and becomes rather ineffective at large tension. For illustration, we consider TREK-1, an established tension-sensitive channel~\cite{brohawn2014mechanosensitivity} that nevertheless associates with the cytoskeleton~\cite{lauritzen2005cross,cox2016removal}. TREK-1 changes its conical angle $\alpha_c \simeq -0.2$ during gating as $\Delta\alpha\simeq 0.36$ while crossing a molecular energy barrier of $[4-7.7]\,\kT$~\cite{maksaev2011analyses}. Gating occurs at tensions in the range of $[0.5-3]\,\kT/\nm^2$. Using Eq.~(\ref{eq_Overall_energy_simpl}), we estimate that a cytoskeletal force of $|F|=10\,\pN$ in the presence of gating-tension leads to energy changes $\sim2\,\kT$, which is smaller than the energy barrier. However, for weak tension, $\gamma \sim 0.01\,\kT/\nm^2$, deformation under force releases $\sim 7\,\kT$, which is clearly comparable to the gating energy barrier. We conclude that some tethered channels likely play a double role as force- and membrane-tension sensors.
\begin{acknowledgments}
We thank the NSF for support via grant MCB-1330288 (to Z.~Gitai and H.A.S.) and
the DAAD for a postdoctoral fellowship (to B.S.).
\end{acknowledgments}


%

\begin{widetext}
\beginsupplement
\newpage
\centering
\vspace*{0.5cm}
\begin{Large}
\textbf{Supplemental Material for ``Role of the Membrane for Mechanosensing by Tethered Channels''} 
\end{Large}
\vspace*{0.5cm}
\flushleft

\section{Membrane theory}
The theory of membranes is very well developed. Technicalities that are necessary to derive the results in the main text
are widely known and can be found in the cited literature. In this supplemental material we intend to summarize the known material and provide step-by-step derivations. 

\subsection{Free energy of a lipid bilayer}
Throughout this work we assume that the membrane-channel systems are in mechanical equilibrium. Physiological processes producing force, such as touch, cell growth, or cell motion occur on timescales of seconds, minutes, or even hours. On the other hand, mechanosensitive membrane channels roughly respond on time scales around $1$~ms to a few $100$~ms. Therefore, time scales are assumed to be well separated.

We consider a lipid bilayer in a Monge representation parametrized by a two-dimensional position vector $\br$ in a reference plane. See Fig.~1 of the main text for a graphical representation of the model. The vertical positions of the upper and lower layers are denoted by $h_{+}(\br)$ and $h_{-}(\br)$ respectively. The equilibrium thickness of the membrane is denoted by $2 \elleq$. For small deviations from the planar case, the membrane energy $\calH$ can be written as
\begin{align}
\calH = \int{\frac{\kappab'}{2}(\nabla^2 h_{+})^2+\frac{\kappab'}{2}(\nabla^2 h_{-})^2 + \elleq K' \left(\frac{h_{+}-h_{-}-2\elleq}{2 \,\elleq}\right)^2+
\gamma'\left(1+\frac{(\nabla h_{+})^2}{2}\right)+\gamma'\left(1+ \frac{(\nabla h_{-})^2}{2}\right)\rmd^2r},
\end{align}
where the surface integration extends over the entire membrane.
The first two terms result from splay-distortion of the leaflets with the single-leaflet bending modulus $\kappab'$. The third term represents the elastic energy due to compression of the membrane in the vertical coordinate. The last two terms represent the area-conservation constraint with a single-leaflet surface tension $\gamma'$. Since we assume small gradients the functional determinant is here approximated as $\sqrt{1+(\nabla h_{\pm})^2} \approx 1+\frac{1}{2}(\nabla h_{\pm})^2$. We now switch to new variables, namely half the equilibrium thickness deviation of the membrane (or thickness deviation of one leaflet) $u$, and the average height $h$, which are defined as 
\begin{subequations}
\begin{align}
u &\equiv (h_{+} - h_{-}-2\elleq)/2,\\
h &\equiv (h_{+} + h_{-})/2.
\end{align}
\end{subequations}
We can split $\calH$ into three contributions to separate all dependencies on $u$ into $\calH_u$ and all dependencies on $h$ into $\calH_h$ as
\begin{align}
\calH =\calH_u +\calH_h +\gamma\int{\rmd^2r},\label{eq_G_full}
\end{align}
with
\begin{subequations}
\begin{align}
\calH_u &= \int{\frac{\kappab}{2}(\nabla^2 u)^2 + \frac{\kappaa}{2} \left(\frac{u}{\elleq}\right)^2+
\frac{\gamma}{2}(\nabla u)^2\rmd^2r},\label{eq_Hu}\\
\calH_h &=\int{\frac{\kappab}{2}(\nabla^2 h)^2 +\frac{\gamma}{2}(\nabla h)^2\rmd^2r},\label{eq_Hh}
\end{align}
\end{subequations}
where we defined $\kappab \equiv 2\kappab'$, $\kappaa \equiv 2 \elleq K'$, and $\gamma \equiv 2\gamma'$. Note that some authors also include a term $\sim \gamma u$, which couples tension linearly to local thickness variation in Eq.~(\ref{eq_Hu})~\cite{ursell2008role}.
Such a term is not considered in this work. 
\subsection{Physiological parameter values}
Typical values of the constants are given in the main text. We usually assume $\gamma = 10^{-3} \,\kT/nm^2$, $\kappab = 25\,\kT$, $2\elleq = 2.85\,\nm$, $\kappaa = 40\,\kT/\nm^2$.
\subsection{Governing equations}
Variation of the energy~(\ref{eq_G_full}) proceeds along common lines by assuming fixed boundaries and using the Euler-Langrange equations for $\calH =  \int{H\rmd^2r}$. When using cartesian coordinates for the plane $x_1,\,x_2$ we have 
\begin{subequations}
\begin{align}
0=&\frac{\p H}{\p u} -\sum_i \frac{\p}{\p x_i} \frac{\p H }{\p(\p u/\p x_i)}+ \sum_i \frac{\p^2}{\p^2 x_i} \frac{\p H }{\p(\p^2 u/\p^2 x_i)}= \kappaa \frac{u}{\ell^2} -\gamma\nabla^2 u +\kappab\nabla^4 u\label{eq_ugoveq}\\
0=&\frac{\p H}{\p h} -\sum_i \frac{\p}{\p x_i} \frac{\p H }{\p(\p h/\p x_i)}+ \sum_i \frac{\p^2}{\p^2 x_i} \frac{\p H }{\p(\p^2 h/\p^2 x_i)}= -\gamma\nabla^2 h +\kappab\nabla^4 h\label{eq_hgoveq}.
\end{align}
\end{subequations}
\subsection{Force and torque}
We assume that forces and torques are applied equally to both leaflets but do not affect the boundary condition for thickness deviation 
$u$ at the channel wall since the latter is dominated by the chemical properties of the channel and membrane. Then, variation of the boundary conditions for $h$ yields an expression for the central force $F$ acting normal to the reference plane~\cite{fournier2007stress}. 
We switch to a cylindrical coordinate system with radius $r$ and angular coordinate $\varphi$ on the reference plane below the membrane.
For a circular membrane inclusion with radius $b$ located at the origin of the coordinate system, the force is expressed as
\begin{align}
F= &-\int_{0}^{2 \pi} b\partial_r \left(\gamma h -\kappab \nabla^2 h \right)|_{r = b} \rmd \varphi.\label{eq_Forcedef}
\end{align}
Likewise, the torque on a circular inclusion can be calculated by variation of the boundaries. The condition for vanishing torque leads to~\cite{weikl1998interaction}
\begin{align}
0= &-\int_{0}^{2 \pi} \left(\p_{r}(\gamma h -\kappab \nabla^2 h)b^2 +\kappab b \nabla^2 h \right)|_{r = b} \cos{\varphi}\,\rmd \varphi.\label{eq_Torquedef}
\end{align}
\section{Calculation of membrane energy $\calFm$ for a single channel}
As seen from Eqns.~(\ref{eq_ugoveq},\ref{eq_hgoveq}), $u$ and $h$ are not directly coupled. Therefore, the calculation of $\calH$ splits into two separate problems for $\calH_u$ and $\calH_h$. 
\subsection{Energy of thickness variation $\calH_u$}
We follow~\cite{nielsen1998energetics} for the calculation of membrane thickness $u$. On dividing Eq.~(\ref{eq_ugoveq}) by $\kappab$ we obtain
\begin{align}
0= \frac{\kappaa}{\kappab\elleq^2}u -\frac{\gamma}{\kappab}\nabla^2 u +\nabla^4 u.\label{eq_ugoveq2}
\end{align}
Rewriting this quadratic form as
\begin{align}
0=(\nabla^2 -\eta^2_{-})(\nabla^2 -\eta^2_{+})u,\label{eq_ugoveq3}
\end{align}
results in
\begin{align}
\eta^2_{\pm} = \frac{1}{2}\left(\frac{\gamma}{\kappab} \pm \sqrt{\left(\frac{\gamma}{\kappab}\right)^2 - \frac{4\kappaa}{\elleq^2 \kappab}}\right).\label{eq_eta}
\end{align}
For the outer boundary of the system at $r=L$ with $L \gg b$ we employ the conditions
\begin{subequations}
\begin{align}
\partial_r u|_{L} &=0,\\
u(L) &=0.
\end{align}
\end{subequations}
For the boundary conditions at the inclusion $r=b$ we use a fixed membrane thickness that is prescribed by 
the height of the hydrophobic channel region $2\ell_{c}$. The difference between the individual leaflet contact angles is denoted by $2\alpha_u$. We have
\begin{subequations}
\begin{align}
u(b) &=\left(\frac{h_{+}-h_{-}-2\elleq}{2}\right)|_{b} =(\ell_{c} - \elleq),\\
\partial_r u|_{b} &= \frac{1}{2}\left(\frac{\partial h_{+}}{\partial r} - \frac{\partial h_{-}}{\partial r}\right)|_{b} =\alpha_u.
\end{align}
\end{subequations}
Solving Eq.~(\ref{eq_ugoveq3}) requires to first solve the two homogeneous equations $(\nabla^2 -\eta^2_{-})u_{hom-}=0$, $(\nabla^2 -\eta^2_{+})u_{hom+}=0$ and then solve the corresponding inhomogeneous equations $(\nabla^2 -\eta^2_{-})u_{in-}=u_{hom+}$, and $(\nabla^2 -\eta^2_{+})u_{in+}=u_{hom-}$. The homogeneous solutions that stay finite for $L\rightarrow \infty$ are given by $A_0^{+} K_0(\eta_{+}r)$ and $A_0^{-} K_0(\eta_{-}r)$ where $K_{n}(x)$ are modified Bessel functions of the second kind. On using these functions for the inhomogeneous equations, it turns out that the inhomogeneous solutions diverge in the large $L$ limit. Thus,
\begin{align}
u(r) = A_0^{+} K_0(\eta_{+} r)+A_0^{-} K_0(\eta_{-}r). 
\end{align}
The constants $A_0^{\pm}$ follow from the boundary conditions as
\begin{align}
A_0^{\pm}=\pm\frac{\alpha_u\, K_0( b \eta_{\mp}) + (\ell_C-\elleq)\,\eta_{\mp}K_1( b \eta_{\mp})}{
\eta_{-} K_0( b \eta_{+}) K_1( b \eta_{-}) - 
 \eta_{+} K_0(b \eta_{-}) K_1( b \eta_{+})}.
\end{align}
The energy contribution due to thickness results from Eq.~(\ref{eq_G_full}) and becomes
\begin{align}
\begin{split}
\calH_u &= b \kappab \pi\frac{\alpha_u K_0(\eta_{-} b) \left[\alpha_u (\eta_{-}^2 - \eta_{+}^2) K_0(\eta_{+} b) + 
        2 \eta_{-}^2 \eta_{+} (\ell_C-\elleq) K_1(\eta_{+} b)\right]}{\eta_{-} K_0(\eta_{+} b) K_1(\eta_{-} b) - \eta_{+} K_0(\eta_{-} b) K_1(\eta_{+} b)} \\
				&+ b \kappab \pi\frac{\eta_{-} \eta_{+} (\ell_C-\elleq) K_1(\eta_{-} b) \left[-2 \alpha_u \eta_{+} K_0(\eta_{+} b) + (\eta_{-}^2 - \eta_{+}^2) (\ell_C-\elleq) K_1(\eta_{+} b)\right]}{\eta_{-} K_0(\eta_{+} b) K_1(\eta_{-} b) - \eta_{+} K_0(\eta_{-} b) K_1(\eta_{+} b)}.\label{eq_Hu1}
\end{split}
\end{align}
\begin{figure}[htb]
  \centering
	\includegraphics[scale=0.53]{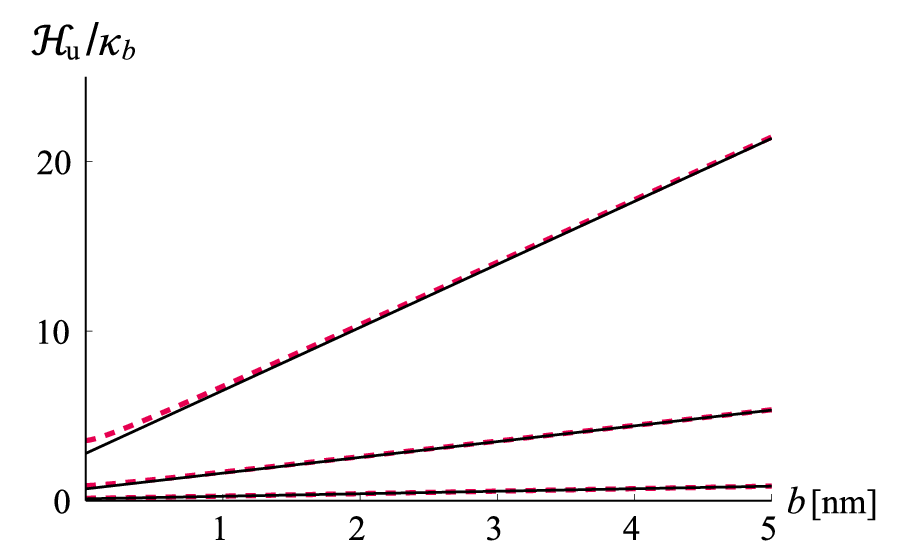} 
\caption{Thickness variation energy $\calH_u$ calculated from Eq.~(\ref{eq_Hu}) and comparison with the approximation Eq.~(\ref{eq_missmatch_Hsupersimpl}). Parameters: $\alpha_u=0$, $\gamma = 10^{-3} \,\kT/\nm^2$, $\kappab = 25\,\kT$, $2\elleq = 3.5\,\nm$, $\kappaa = 40\,\kT/\nm^2$.} 
\label{fig_approx_cmp}
\end{figure}
This lengthy expression can be simplified by assuming $\gamma^2 \ll 4 \kappab \kappaa/\elleq^2$, which is appropriate for the physiological range of constants given above. Then, Eq.~(\ref{eq_eta}) approximated by 
\begin{align}
\eta^2_{\pm}\approx \gamma/(2 \kappab)\pm i \left(\sqrt{\kappaa/\kappab}\right)/\elleq\label{eq_approx_eta}.
\end{align}
The lengthscale associated with $\eta_{\pm}$ is then given by
\begin{equation}
(\eta_{+}^{-1}+\eta_{-}^{-1}) \approx  (4\kappa_{\rm{b}} \elleq^2/\kappa_{\rm{a}})^{1/4} \simeq 1.5\,\nm,
\end{equation}
where we employed the parameter values given above.
Since the channel radius is usually larger than the lengthscale of tickness perturbations $(\eta_{+}^{-1}+\eta_{-}^{-1})<b$ we expand $\calH_u$ for large $b$ to obtain
\begin{equation}
\begin{split}
\calH_u &=  \frac{\kappab \pi}{2} (2 \alpha_u (\eta_{-} + \eta_{+}) (\ell_{c}-\elleq) + (\eta_{-} + \eta_{+})^2 (\ell_{c}-\elleq)^2) \\
&+ \kappab \pi (\alpha_u^2 + \frac{2 \eta_{-} \eta_{+}}{\eta_{-} + \eta_{+}} \alpha_u (\ell_{c}-\elleq) + \eta_{-} \eta_{+}  (\ell_{c}-\elleq)^2) (\eta_{-} + 
    \eta_{+}) b\\ 
		&- \frac{
 \kappab \pi (\eta_{-} + \eta_{+}) (\alpha_u^2 + 
    2 \alpha_u (\eta_{-} + \eta_{+}) (\ell_{c}-\elleq) + (\eta_{-}^2 - \eta_{-} \eta_{+} + \eta_{+}^2) (\ell_{c}-\elleq)^2)}{8 \eta_{-} \eta_{+} b} +O\left(\left[\frac{\eta_{-}+\eta_{+}}{\eta_{-} \eta_{+} b}\right]^2\right).
	\label{eq_missmatch_Hsimp}
\end{split}
\end{equation}
To simplify the equation even more, we assume that $\alpha_u=0$ ~\cite{huang1986deformation}. Such an assumption is reasonable for membrane proteins that have a straight outer wall which imposes the same contact angle on the two leaflets. Employing now 
the parameter values given in the main text, we find that $\calH_u$ can be well approximated well by a linear relationship with positive slope
\begin{equation}
\calH_{u}\approx \frac{\kappab \pi}{2}(\eta_{-} + \eta_{+})^2(\ell_{c}-\elleq)^2 + \pi \kappa_{\rm{b}}(\ell_{c}-\ell_{\rm{eq}})^2 \eta_{+}\eta_{-}(\eta_{+}+\eta_{-}) b,\label{eq_missmatch_Hsupersimpl}
\end{equation}
which is the simplified expression given in the main text. Fig.~(\ref{fig_approx_cmp}) shows a comparison of the approximation Eq.~(\ref{eq_missmatch_Hsupersimpl}) and the full expression Eq.~(\ref{eq_Hu1}). For a channel radius $b > 1.5\,\nm$, the linear approximation is seen to hold quite well.  
\subsection{Energy of height variation $\calH_h$}
The free membrane shape is determined by Eq.~(\ref{eq_hgoveq}), which is written as
\begin{align}
\nabla^2(\nabla^2 -\xi^2)h =0,\label{eq_shape_fluid_mem}
\end{align}
with $\xi^2 = \gamma/\kappab$. The solution is to be radially symmetric, with fixed boundary conditions 
at outer boundary of the system $r=L$ as
\begin{subequations}
\begin{align}
h(L) &= 0,\\
\p_r h(r)|_{r=L} &= 0.
\end{align}
\end{subequations}
At the channel wall with radius $b$ the average contact angles of both leaflets is denoted by $\alpha$. 
We assume that $\alpha$ is fixed and dictated by the shape of the channel since lipid molecules tend to 
align side-by-side with the channel walls. The boundary conditions at the protein are thus given by
\begin{subequations}
\begin{align}
h(b) &= h(0) =\text{const.},\\
\p_r h(r)|_{r=b} &= \alpha.
\end{align}
\end{subequations}
The final boundary condition determining $h(0)$ is the force balance at the protein center
where a constant vertical force $F$ is applied
\begin{align}
F= &-\kappab 2 \pi \p_{r}\left(\xi^2 h -\nabla^2 h \right) r|_{r = b}.\label{eq_FF1}
\end{align}
Then, the solution of Eq.~(\ref{eq_shape_fluid_mem}) is given by 
\begin{align}
h = \frac{F\log(L/r)}{2 \gamma \pi} -\frac{(F + 2 b \alpha \gamma \pi) K_0(r \xi)}{2 b \gamma \pi \xi K_1(b \xi)},
\end{align}
where we have dropped all terms that decay exponentially with $\xi L$. Next, we calculate $\calH_h$ from Eq.~(\ref{eq_G_full}). 
The gradient term can be expressed as $(\nabla h)^2=\nabla\cdot(h\nabla h)-h\nabla^2h$. Using the governing equations, we find
\begin{align}
\begin{split}
\calH_{h} =& -\frac{\gamma}{2}\int h \p_r h \,r\rmd \varphi_b|_{r = b}+\frac{\gamma}{2}\int h \p_r h \,r\rmd \varphi_b|_{r = L}-\frac{\gamma}{2}\int h^{H} \nabla^2h^{S} \rmd^2 r,
\end{split}
\end{align}
where $h^{H}$ and $h^{S}$ are the parts of $h$ that fulfill $\nabla^2h^{H}=0$ and $(\nabla^2+\xi^2)h^{S}=0$.
Again, we ignore terms that decay exponentially with $L$ and obtain up to a constant
\begin{align}
\calH_{h}= -\frac{F^2 K_0(b\xi)}{4 b \gamma \pi \xi K_1(b\xi)}+\frac{F^2 \log(L/b)}{4 \gamma \pi}+
\frac{  (2 b \gamma \pi  \alpha)^2 K_0(b \xi)}{4 b \gamma \pi \xi K_1(b \xi)}.
\end{align}
\subsection{Overall membrane energy}
One part of the overall membrane energy is given by the energy related to thickness and height perturbations $\calH=\calH_u+\calH_h$. Since a size change of the channel affects the membrane area, the term $\gamma\int{\rmd^2r}$ in Eq.~(\ref{eq_G_full}) produces a second contribution to the overall energy. Third, we must add a term $-\int_0^{h} F\,\rmd h'$ to take into account the work done by the the external force $F$. The result is
\begin{equation}
\calFm =\calH_u +\calH_h +\gamma\int{\rmd^2r} -\int_0^{h} F\,\rmd h'.
\end{equation}
Assuming constant force applied on the channel at $r=0$, we obtain for the work 
\begin{align} 
-F h(0)=-F h(b) = -\frac{F^2\log(L/b)}{2 \gamma \pi} +\frac{(F^2 + 2 b \gamma \pi \alpha F ) K_0(b \xi)}{2 b \gamma \pi \xi K_1(b \xi)}.\label{eq_elast_term}
\end{align}
Adding $\calH_h$ yields 
\begin{align}
\calH_{h} - F h(0) = -\frac{F^2\log(L/b)}{4\gamma \pi} +\frac{(F + 2 b \gamma \pi \alpha F )^2 K_0(b \xi)}{4 b \gamma \pi \xi K_1(b \xi)}.\label{eq_energy_h}
\end{align}
To finally calculate the overall energy, we use $\gamma \int\rmd r^2 = \gamma \pi(L^2-b^2) \sim -\gamma \pi b^2$ and obtain
\begin{align}
\begin{split}
&\calFm=\calH_{u}-\gamma \pi b^2 +\frac{(F + 2\pi\gamma b \alpha )^2}{4 \pi \gamma} \frac{K_0(b\xi)}{b\xi K_1(b\xi)}-\frac{F^2\,\log(L/b)}{4 \pi \gamma},\label{eq_Overall_free_energy}
\end{split}
\end{align}
The last term in Eq.~(\ref{eq_Overall_free_energy}) displays a logarithmic divergence with system size $L$, which is immaterial since we consider only fixed $L$ and $F$. Such energy terms that do not depend on channel shape parameters are removed by defining $\tcalFm \equiv \calFm + F^2\,[\Gamma_e +\log(\xi L/2)]/(4 \pi \gamma)$, where $\Gamma_e=0.577\ldots$ is the Euler-Mascheroni constant. The equation for $\tcalFm$ in the main text is an expansion of Eq.~(\ref{eq_Overall_free_energy}) for $\xi b\ll 1$.
\section{Model extensions}
\subsection{Elastic interaction between membrane and environment}
Sub-membrane structures can damp deformations, which possibly affects the energetics of channel shape change. We model a supported membrane in a mean-field approach by adding a quadratic potential with stiffness per area $\epsilon$ to Eq.~(\ref{eq_Hh}) as $\calH_{h,\epsilon} = \calH_h + \int{\frac{\epsilon}{2}h^2\rmd^2r}$. In analogy to the equations for thickness perturbation $u$, we now have two scales for height $\nu^2_{\pm} \equiv (\xi^2 \pm \sqrt{\xi^4 - 4\epsilon/\kappab})/2$.  We obtain for $\xi^4 < 4\epsilon/\kappab$ the expression 
\begin{equation}
\calF_{m,\epsilon} = \calH_{u} -\gamma \pi b^2+ \left[\frac{(F + 2\pi b \kappab \nu_{+}^2\alpha)^2 K_0(b\nu_{+})}{4 \pi \kappab b (\nu_+^2 -\nu_{-}^2) \nu_{+} K_1(b\nu_{+})} + \mathrm{c.c.}\right],\label{scheiss1}
\end{equation}
which has a similar form as Eq.~(\ref{eq_Overall_free_energy}). In analogy to the effect of increasing membrane tension shown in Fig.~2a) of the main text, increasing $\epsilon$ reduces the energy that can be gained by conical deformation. The membrane lever that amplifies externally applied forces leading to small thresholds $F^{*} < \delta_{\alpha}/b$ becomes ineffective when $\sqrt[4]{\kappab/\epsilon}\lesssim b$, which typically occurs for $\epsilon \gtrsim 0.1\,\pN/\nm^3$.

If the effect of the bare membrane tension is much smaller than the effect of the elastic support $\gamma^2 \ll 4 \epsilon \kappab$, a single lengthscale $\xi_{\epsilon}^{-1}  =\sqrt[4]{\kappab/\epsilon}$ governs membrane deformation. When expanding Eq.~(\ref{scheiss1}) for $\xi_{\epsilon} b\ll 1$ we obtain
\begin{equation}
\calF_{m,\epsilon} \approx \calH_{u} - \gamma b^2 \pi  + 
   b \alpha F (-\Gamma_e - \log(\frac{b \xi_{\epsilon}}{2})) - \frac{F^2}{16 \xi_{\epsilon}^2 \kappab}+O(b^2\xi_{\epsilon}^2),
\end{equation}
which has a similar form as Eq.~(5) of the main text.
\subsection{Effect of lipid orientation on membrane energy}
The mesoscopic description for lipid bilayers employed in the main text can be modified to
account for the microscopic orientational degree of freedom of lipid molecules~\cite{mackintosh1991orientational,seifert1996role,fournier1999microscopic, kozlovsky2004orientation,may2004tilt,argudo2016continuum}. Spatial variations of 
lipid orientation in the leaflets produce an elastic energy. It has been demonstrated in experiments that tilt can affect membrane deformations on a lengthscale smaller than a few bilayer thicknesses~\cite{jablin2014experimental}. However, 
the microscopic details of membranes around channels are to date somewhat unclear, and 
even molecular-scale simulations do not necessarily reproduce the crowded environment of a real membrane correctly.
Nevertheless, depending on lipid composition, lipid tilt may be relevant for opening of membrane channels. 
To assess the role of lipid tilt for the force sensing mechanisms discussed here, we distinguish between membrane thickness 
deviation $u$ and average membrane height $h$.

Membrane thickness deviations $u$ occur on short lengthscales $(\eta_{+}^{-1}+\eta_{-}^{-1}) \sim 1\,\nm$, where the lipid tilt could most likely affect membrane shape. However, $u$ only plays a role for the force-independent term $\sim d^2 b$ in the central equation~(5) 
of the main text and is not pivotal for the discussed force-sensing mechanisms. The lumped expression $d^2$ also depends on an unknown change of the hydrophobic thickness of the channel during opening, on microscopic distribution of hydrophobic residues on the channel molecule, and on the characteristics of the membrane. Including the effect of lipid tilt does not necessarily lead to a more quantitative prediction of $d^2$ and we do not attempt such a description here. The reader is referred to~\cite{fournier1999microscopic, may2002membrane,kuzmin2005line, venturoli2006mesoscopic} for further discussions of the effect of lipid tilt on membrane thickness.

In contrast to $u$, the height $h$ of the membrane midplane couples to force $F$. To quantify the influence of lipid tilt on the energy of midplane deformation, we extend the frameworks used in Refs.~\cite{fournier1999microscopic, kozlovsky2004orientation, watson2013thermal} to the case of finite force $F$. The normal vectors at the upper ($+$) and lower ($-$) leaflets pointing towards the midplane are denoted by $\bN^{\pm} \approx \mp\hat{\mathbf{e}}_z \pm \hat{\mathbf{e}}_r\p_r h^{\pm}(r)$ (Fig.~\ref{fig_tiltskech}). The orientation of the lipids is described by nematic directors $\bt^{\pm}$, which are the unit vectors pointing along the center of the hydrocarbon chains towards the center of the bilayer. Assuming that the directors are almost parallel to the leaflet normals, lipid tilt is written as
\begin{align}
\bmm^{\pm} = \frac{\bt^{\pm}}{\bt^{\pm} \bN^{\pm}} -\bN^{\pm} \approx \bt^{\pm} -\bN^{\pm}.
\end{align}
Then, the average tilt vector $\bmm$ and the tilt-difference vector $\bmd$ are defined as
\begin{align}
\bmm \equiv \frac{\bmm^{+}-\bmm^{-}}{2}, & & \bmd \equiv \frac{\bmm^{+}+\bmm^{-}}{2}.\label{eq_orient}
\end{align}
For the boundary condition along the channel wall we assume that the average orientation of the lipids is parallel to the channel, which means $-(\bt^{+}-\bt^{-})/2|_{r=b}\approx-\alpha\hat{\mathbf{e}}_r+\hat{\mathbf{e}}_z$. Using Eq.~(\ref{eq_orient}) along with the definition of the midplane height $h=(h^{+}+h^{-})/2$, we find 
\begin{align}
\begin{split}
\alpha \approx \hat{\mathbf{e}}_r\,(\nabla h + \bmm)|_{r=b},\\
0 =  \hat{\mathbf{e}}_z\,\bmm|_{r=b}.
\end{split}\label{eq_bcchanneltilt}
\end{align}
Note that this condition on the average tilt does not imply that the lipid tilt in individual leaflets vanishes. 
Interactions between hydrocarbon chains and the channel wall can possibly lead to $\bmd|_{r=b}\neq \mathbf{0}$. We also assume that all tilt vanishes at the periphery of the membrane, which is at $r=L$ far away from the channel
\begin{equation}
\bmm|_{r=L} = \mathbf{0}.
\end{equation}

\begin{figure}[htb]
  \centering
	\includegraphics[scale=1]{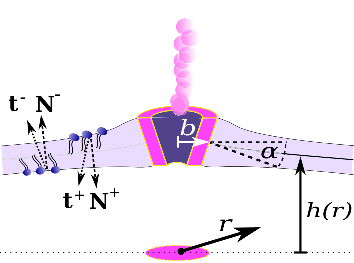} 
\caption{Sketch of the channel model with leaflet normals $\bN^{\pm}$ and a lipid orientation that is described by the directors $\bt^{\pm}$.} 
\label{fig_tiltskech}
\end{figure}

When considering lipid orientation, the energy expression related to membrane height needs to be complemented by terms for the elastic energy stored in tilt magnitude and spatial variations of tilt. The standard formulation including splay $\sim\nabla\cdot\bmm$ and twist $\sim\nabla\times\bmm$ of the director field is
\begin{align}
\calH_{h}^{\rm{tilt}} =\frac{1}{2}\int{\left[\kappab(\nabla^2 h +\nabla\cdot\bmm)^2 +\kappasp \bmm^2 +\kappatw (\nabla \times \bmm)^2 + \gamma(\nabla h)^2 +\tilde{\gamma}\right]\rmd^2r},\label{eq_Hht}
\end{align}
where $\tilde{\gamma}$ is a new constant combining the effect of tilt and surface tension on membrane area.
The new elastic constants $\kappasp$ and $\kappatw$ are related to tilt magnitude and twist modes in the director field.
The splay mode was absorbed in the curvature term with bending constant $\kappab$~\cite{jablin2014experimental}. As seen from Eqns.~(\ref{eq_Hht}), the mean tilt $\bmm$ couples to $h$, while the tilt difference $\bmd$ does not affect $h$.

The Euler equations are derived as usual through variation of Eq.~(\ref{eq_Hht}). The requirements $\frac{\delta \calH_h^{\rm{tilt}}}{\delta h} =0$ and $\frac{\delta \calH_h^{\rm{tilt}}}{\delta \bmm} =0$ in the interior of the membrane area yield
\begin{subequations}
\begin{align}
&\kappab  \nabla^2(\nabla^2h +\nabla\cdot\bmm) -\gamma\nabla^2h = 0,\label{eq_euh}\\
&\kappasp \bmm -\kappab \nabla(\nabla^2h +\nabla\cdot\bmm) +\kappatw \nabla\times(\nabla\times \bmm) = 0.\label{eq_eum}
\end{align}
\end{subequations}
Taking the divergence of Eq.~(\ref{eq_eum}) and insertion of the result into Eq.~(\ref{eq_euh}) yields  
\begin{align}
\nabla\cdot\bmm = \frac{\gamma}{\kappasp}\nabla^2h.\label{eq_tuttta}
\end{align}
With Eq.~(\ref{eq_tuttta}), the Euler equations become
\begin{subequations}
\begin{align}
&\tkappab  \nabla^2( \nabla^2 h -\frac{\gamma}{\tkappab}h) = 0,\label{eq_euh2}\\
&-\kappatw \nabla^2 \bmm + \kappasp \bmm +(\frac{\kappatw\gamma}{\kappasp}-\tkappab)\nabla(\nabla^2h)=0,\label{eq_eum2}
\end{align}
\end{subequations}
where we defined a rescaled bending constant as
\begin{equation}
\tkappab \equiv \kappab \left(1 +\frac{\gamma}{\kappasp}\right).\label{eq_rescalekappa}
\end{equation}
The lengthscale governing height variations in the presence of tilt is thus given by
\begin{equation}
\txi^{-1} \equiv \sqrt{\tkappab/\gamma}.\label{eq_rescalexi}
\end{equation}
Note that Eq.~(\ref{eq_eum2}) can be simplified even more by applying a curl and using the fact that the curl commutes with the 
Laplacian of a vector field. We obtain
\begin{align}
-\kappatw \nabla^2 (\nabla\times\bmm) + \kappasp (\nabla\times\bmm)=0.\label{eq_eum3}
\end{align}
To derive the force balance, we consider variation of the boundary conditions in Eq.~(\ref{eq_Hht}). The director field $\bmm$ can be written as sum of two independent fields $\bmm = \bmm^{\rm{p}} +\bmm^{\rm{c}}$ where $\nabla\times \bmm^{\rm{p}}=0$ and $\nabla\cdot\bmm^{\rm{c}}=0$. Variation of $\calH_h^{\rm{tilt}}$ with respect to $h$ and $\bmm$ and use of Eqns.~(\ref{eq_euh},\ref{eq_eum}) yields
\begin{align}
\delta \calH_h^{\rm{tilt}}|_{bc} = \int [ \kappab (\nabla^2 h +\nabla\cdot\bmm^{\rm{p}})(\nabla \delta h +\delta\bmm)-\kappab \nabla(\nabla^2 h -\frac{\gamma}{\kappab} h + \nabla\cdot\bmm^{\rm{p}})\delta h + \kappatw \delta\bmm \times (\nabla\times \bmm^{\rm{c}})]\mathbf{n} \rmd s,\label{eq_var_bound}
\end{align}
where $\bn$ denotes a normal vector in the membrane plane pointing outwards from the membrane area determined by the contour path $s$. Note that the usage of nabla operator rules to derive Eq.~(\ref{eq_var_bound}) yielded a variation $\delta\bmm$ that was not split into $\bmm^{\rm{p}}$ and $\bmm^{\rm{c}}$ as in the terms $\nabla\cdot\bmm^{\rm{p}}$ and $\nabla\times\bmm^{\rm{c}}$. Due to the boundary conditions at the channel walls~(\ref{eq_bcchanneltilt}), $(\nabla \delta h+\delta\bmm)|_{r=b}$ must vanish. Therefore, tilt and height variations are not independent and the first term in Eq.~(\ref{eq_var_bound}) vanishes. Furthermore, we assume $(\nabla\times \bmm^{\rm{c}})|_{r=b} = 0$ to have a vanishing last term. We are thus left with the condition
\begin{align}
0=\frac{\delta \calH_h^{\rm{tilt}}|_{bc}}{\delta h} = \int [ -\kappab \nabla(\nabla^2 h -\frac{\gamma}{\kappab} h + \nabla\cdot\bmm^{\rm{p}})]\mathbf{n} \rmd s= \int -\tkappab \nabla(\nabla^2 h -\txi^2 h)\mathbf{n} \rmd s,\label{eq_var_bound2}
\end{align}
which expresses a balance of all forces applied to the membrane edges. In our case, the force $F$ on the channel is balanced by a counterforce acting on the membrane perimeter at $r=L$. From Eq.~(\ref{eq_var_bound2}), we see that the formula for force $F$ is exactly the same as in absence of tilt (Eq.(\ref{eq_FF1})) when $\kappab$ is replaced with the rescaled quantity $\tkappab$ as 
\begin{align}
F= &-\tkappab 2 \pi \p_{r}\left(\txi^2 h -\nabla^2 h \right) r|_{r = b}.\label{eq_FF2}
\end{align}
The solutions that satisfy the imposed boundary conditions are
\begin{subequations}
\begin{align}
h &= \frac{F\log(L/r)}{2 \gamma \pi} -\frac{(F + 2 b \alpha \gamma \pi) K_0(r \txi)}{2 b \pi \gamma (1 + \gamma/\kappasp) \txi K_1(b \txi)},\\
\bmm &= \hat{\mathbf{e}}_r\,\frac{(F + 2 b \alpha \gamma \pi) K_0(r \txi)}{2 b \pi  (\kappasp + \gamma)  K_1(b \txi)}.
\end{align}
\end{subequations}
We insert these solutions into Eq.~(\ref{eq_Hht}) and then calculate the energy in the constant force ensemble 
\begin{align}
\begin{split}
&\calFm^{\mathrm{tilt}}=\calH_{u}^{\mathrm{tilt}}-\tgamma \pi b^2 +\frac{(F + 2\pi\gamma b \alpha )^2}{4 \pi \gamma\,(1 + \gamma/\kappasp)} \frac{K_0(b\txi)}{b\txi K_1(b\txi)}-\frac{F^2\,\log(L/b)}{4 \pi \gamma}.\label{eq_Overall_free_energy_tilt}
\end{split}
\end{align}
This expression has the same form as the analogous formula for the model without lipid tilt, Eq.~(\ref{eq_Overall_free_energy}). The difference occurs through a rescaling of the bending constant given in Eqns.~(\ref{eq_rescalekappa},\ref{eq_rescalexi}). Measurements~\cite{jablin2014experimental} and simulations~\cite{watson2013thermal} have reported that $\kappasp \simeq [10-25]\,\kT/\nm$. For unusually large membrane tension we have at most $\gamma/\kappasp \sim 0.1$. Therefore, the effect of tilt on midplane deformation energy is small.
\section{Experiments: Membrane thickness perturbation and gating probability} 
Controlled mechanical stimulation of membrane channels can be done either by a patch-clamp measurements~\cite{prager2014unique,zhang2015ankyrin} or by applying force to the membrane with a bead in an optical trap~\cite{hayakawa2008actin}. The latter technique has the advantage that it does not perturb the membrane tension strongly, which allows investigation of the tether mechanism. 
Optical traps and magnetic beads allow to apply constant forces, which corresponds to the constant force ensemble
studied in this letter. Measurement of the channel response requires here either a whole-cell patch-clamp technique or 
fluorescent imaging of the transported ion~\cite{hayakawa2008actin}.

\begin{figure}[htb]
  \centering
	\includegraphics[width=0.83\linewidth]{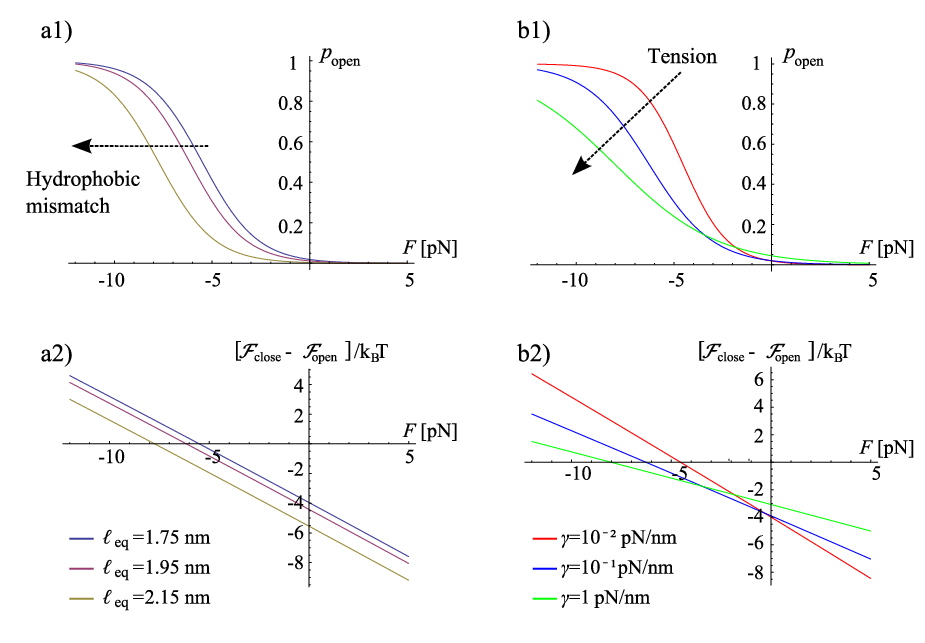} 
\caption{Gating curves for a force-sensing two-state channel. a1)~Increasing hydrophobic mismatch shifts the occurrence of gating to higher absolute force values. a2)~The energy difference between open and closed state is shifted while maintaining the same slope.
b1)~Increasing membrane tension also increases the absolute force necessary for gating. b2) Tension changes the slope of the force-dependent energy difference between the states.} 
\label{fig_gating_p}
\end{figure}

If the gating probability along with applied forces can be measured, a variation of membrane thickness and tension allows to investigate  membrane-dependent gating mechanisms. For a two-state channel in equilibrium, the gating probabilities at constant force are given by
\begin{align}
p_{\mathrm{open}} = e^{-\frac{\calF_{\mathrm{open}}}{\kT}}/\left(e^{-\frac{\calF_{\mathrm{open}}}{\kT}}+e^{-\frac{\calF_{\mathrm{close}}}{\kT}}\right).
\end{align}
Measurement of probabilities also allows to reconstruct the energy difference between open and closed states as
\begin{align}
\log\left(p_{\mathrm{open}}/p_{\mathrm{close}}\right) = \left(-\calF_{\mathrm{open}}+\calF_{\mathrm{close}}\right)/\kT.
\end{align}
Since parameter values for genuine force-sensitive channels such as TRPN are presently not available, we assume channel 
properties that are similar to those of TREK-1~\cite{maksaev2011analyses}, summarized in Tab.~\ref{tab:dfdd}. Figure~\ref{fig_gating_p} displays the calculated gating probabilities for the model channel. In normal conditions, gating occurs when the channel is pulled with $\sim-5\,\pN$. If the membrane thickness is changed, e.g., biochemically, a mismatch between the hydrophobic channel thickness and the membrane thickness shifts gating to higher absolute force values. Figure~\ref{fig_gating_p}a2) demonstrates that changing hydrophobic mismatch leads to a parallel translation of the $\log\left(p_{\mathrm{open}}/p_{\mathrm{close}}\right)$ curve. Figure~\ref{fig_gating_p}b1) illustrates the dependence of the gating curve on the membrane tension. High tension increases the absolute values of force needed for gating. However, in contrast to hydrophobic mismatch, tension changes the slope of the $\log\left(p_{\mathrm{open}}/p_{\mathrm{close}}\right)$ curves, as seen in Fig.~\ref{fig_gating_p}b2).
\begin{table*}[h]
	\centering
		\begin{tabular}{ |l | c | r| }
			\hline			
			Conical angle (closed) & $\alpha_c$ & $-0.22$ rad \\
			Angle change for gating & $\Delta \alpha$ & $+0.38$ rad\\
			Radius (closed) & $b_c$ & $2.37\,\nm$ \\
			Radius change for gating & $\Delta b$ & $0.2\,\nm$ \\
			Internal energy (closed) & $\calF_{\rm{int, c}}$ &$-4\,\kT$\\
			Internal energy (open) & $\calF_{\rm{int, o}}$ & $0$\\
			Hydrophobic channel thickness (closed and open) & $2\,\ell_C$ & $2 \times 1.75\,\nm$\\
			Bending constant & $\kappab$& $25\,\kT$\\
			Compression modulus & $\kappaa$& $40\,\kT/\nm^2$\\
			\hline  
		\end{tabular}		
	\caption{Representative parameter values chosen for calculation of gating probabilities in Fig.~\ref{fig_gating_p}}
	\label{tab:dfdd}
\end{table*}

\section{Interaction between two channels}
We consider two circular channels as depicted in Fig.~\ref{fig_two_inclusions}a). The channel radii are denoted by $b_{\{1,2\}}$. 
The center-to-center distance between the proteins is denoted by $R$. Following Ref.~\cite{weikl1998interaction}, we take $R$ to be much larger than $b_1$ and $b_2$ and use a multipole expansion to calculate interaction energies. Since the thickness perturbations $u$ decay on a $\nm$-lengthscale, channel interaction due to $u$ will become negligible in the considered limit and we focus on interaction through height $h$. 

To avoid the mathematical complications arising from long-range deformations, we assume here a supported membrane where vertical membrane displacements $h$ are penalized by a quadratic potential with stiffness per area $\epsilon$. As the very last step we will consider the limit $\epsilon\rightarrow 0$. Employing a rescaled elastic constant $\vartheta \equiv \epsilon/\kappab$ the new Hamiltonian governing $h$ reads
\begin{align}
\calH_{h,\epsilon} = \frac{\kappab}{2}\int\left[(\nabla^2 h)^2+\xi^2(\nabla h)^2 + \vartheta h^2\right]\,\rmd^2 r,\label{eq_Hs_surf}
\end{align}
where the integral extends over the whole membrane. The Euler-Lagrange equations determining the membrane shape are now
\begin{align}
0=\nabla^2(\nabla^2 -\xi^2)h +\vartheta h = \nabla^2\nabla^2h -(\nu^2_{+}+\nu^2_{-})\nabla^2h +\nu^2_{+}\nu^2_{-}h= (\nabla^2 - \nu^2_{+})(\nabla^2 - \nu^2_{-})h,\label{eq_supp_mem}
\end{align}
where we defined two scales
\begin{align}
\nu^2_{\pm} \equiv (\xi^2 \pm \sqrt{\xi^4 - 4\vartheta})/2.\label{eq_nudef}
\end{align}
Note that the $\nu^2_{\pm}$ become imaginary if the elastic constant is large $\xi^2 - 4\vartheta<0$. Then we expect decaying oscillatory solutions for $h$. 

As shown in Fig.~\ref{fig_two_inclusions}a), we employ two polar coordinate systems $(r_k,\varphi_k)$, $k\in\{1,2\}$ that are centered on either of the two channels. Every point on the membrane can be described by any of these systems. Figure~\ref{fig_two_inclusions}b) shows the angles that are used to parametrize the channel-membrane interface. The channels are tilted by angles $\beta_{\{1,2\}}$ with respect to the horizontal plane. The conical angles $\alpha_{\{1,2\}}$ are determined by the shape of the channels. Taken together, the boundary conditions for the membrane at the channels become
\begin{subequations}
\begin{align}
h(r_1, \varphi_1)|_{r_1=b_1} &= h(r_1=0) + b_1\,\beta_1 \cos(\varphi_1),\\
h(r_2, \varphi_2)|_{r_2=b_2} &= h(r_2=0) + b_2\,\beta_2 \cos(\varphi_2),\\
\p_{r_1} h(r_1, \varphi_1)|_{r_1=b_1} &= \alpha_1 + \beta_1 \cos(\varphi_1),\\
\p_{r_2} h(r_2, \varphi_2)|_{r_2=b_2} &= \alpha_2 + \beta_2 \cos(\varphi_2).
\end{align}
\end{subequations}
Vertical forces $F_{\{1,2\}}$ are applied to the center of both channels. These forces result in membrane deformation. 
However, since the channels can tilt freely, the torque on the membrane vanishes. These conditions allow to determine the 
tilt angles $\beta_{\{1,2\}}$ and the height at the channel centers $h(r_{\{1,2\}}=0)$. The corresponding boundary conditions read
\begin{subequations}
\begin{align}
0= &\kappab \int_{0}^{2 \pi} \left[\p_{r_1}(\xi^2 h -\nabla^2 h)b_1 +\nabla^2 h \right] b_1 \cos{\varphi_1}\,\rmd \varphi_1|_{r_1 = b_1},\\
0= &\kappab \int_{0}^{2 \pi} \left[\p_{r_2}(\xi^2 h -\nabla^2 h)b_2 +\nabla^2 h \right] b_2 \cos{\varphi_2}\,\rmd \varphi_2|_{r_2 = b_2},\\
F_1= &\kappab \int_{0}^{2 \pi} \p_{r_1}\left(\nabla^2 h -\xi^2 h\right) b_1 \rmd \varphi_1|_{r_1 = b_1},\\
F_2= &\kappab \int_{0}^{2 \pi} \p_{r_2}\left(\nabla^2 h -\xi^2 h\right) b_2 \rmd \varphi_2|_{r_2 = b_2}.
\end{align}
\end{subequations}
\begin{figure}[htb]
  \centering
	\includegraphics[scale=0.57]{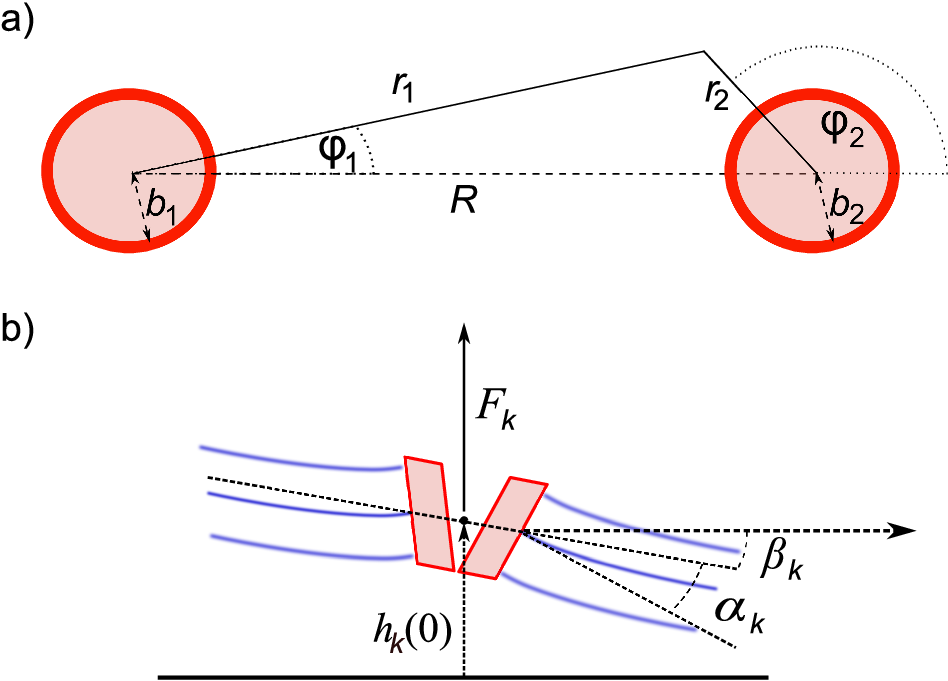} 
\caption{a) Top view on the geometry used to describe interaction between two channels in the membrane. Vertical forces $F_{\{1,2\}}$ are applied on the center of the channels. b) Side view on a channel. $\beta_k$ is the tilt angle of the channel mid-plane with respect to a horizontal plane. $\alpha_k$ is the conical angle between a channel wall normal and the mid-plane. $\alpha_k$ is determined by the channel shape.} 
\label{fig_two_inclusions}
\end{figure}
We aim to calculate $h(r)$ in the close vicinity of either of the channels. Assuming that $R\gg b_{\{1,2\}}$, one might write $h(r)$
close to channel $k$ as solution for only that channel plus a perturbation from the other channel. For one channel only, the solution of Eq.~(\ref{eq_supp_mem}) is given by a linear combination of the solutions of $(\nabla^2-\nu^2_{-})h^{-}_k=0$ and $(\nabla^2-\nu^2_{+})h^{+}_k=0$ as
\begin{align}
h_k &= h^{+}_k + h^{-}_k,\\
h^{\pm}_k &= \sum_{n=0}^{\infty} c_n^{k\pm} K_n(\nu_{\pm} r_k) \cos(n\varphi_k) \approx c_0^{k\pm} K_0(\nu_{\pm} r_k)+c_1^{k\pm} K_1(\nu_{\pm} r_k) \cos(\varphi_k),
\end{align}
where the multipole series is truncated after the first moment. Close to channel $1$, we wish to express the perturbation due to channel $2$ in coordinates $r_1,\varphi_1$. The pertaining geometrical relations (see Fig.~\ref{fig_two_inclusions}), can be expanded in powers of $r_1/R$ as 
\begin{align}
\begin{split}
&r_2 = \sqrt{R^2 + r_1^2 - 2 R r_1 \cos(\varphi_1)}\approx R - r_1 \cos(\varphi_1) + \frac{(1-\cos^2(\varphi_1))\,r_1^2}{2 R},\\
&\cos(\varphi_2) = \left(r_1 \cos(\varphi_1) -R\right)/r_2\approx -1 + \frac{(1 - \cos^2(\varphi_1))\,r_1^2}{2 R^2}.\label{eq_exp_close_to_a}
\end{split}
\end{align}
Conversely, close to channel $2$ we have  
\begin{align}
\begin{split}
&r_1 = \sqrt{R^2 + r_2^2+2 R r_2 \cos(\varphi_2)},\\
&\cos(\varphi_1) = \left(r_2 \cos(\varphi_2) +R\right)/r_1.\label{eq_exp_close_to_b}
\end{split}
\end{align} 
Using these relations, we express $h$ close to either of the two channels. For example, in the vicinity of channel $1$, a Taylor expansion in powers of $r_1/R$ leads to
\begin{align}
\begin{split}
h(r_1,\varphi_1)|_{r_1 \approx b_1} &\approx
c^{1-}_{0} K_{0}(\nu_{-} r_{1}) + c^{1+}_{0} K_{0}(\nu_{+} r_{1}) + 
 \left(c^{1-}_{1} K_{1}(\nu_{-} r_{1}) + c^{1+}_{1} K_{1}(\nu_{+} r_{1})\right) \cos(\varphi_1) \\
&+ c^{2-}_{0} (K_{0}(\nu_{-} R) + r_{1} \nu_{-} K_{1}(\nu_{-} R) \cos(\varphi_1)) + 
 c^{2+}_{0} (K_{0}(\nu_{+} R) + r_{1} \nu_{+} K_{1}(\nu_{+} R) \cos(\varphi_1))\\ 
&+ c^{2-}_{1} (-K_{1}(\nu_{-} R) - \frac{r_{1}^2 \nu_{-}^2}{4} K_{1}(\nu_{-} R) - 
    \frac{r_{1} \nu_{-}}{2} (K_{0}(\nu_{-} R) + K_{2}(\nu_{-} R)) \cos(\varphi_1))\\
&+ c^{2+}_{1} (-K_{1}(\nu_{+} R) - \frac{r_{1}^2 \nu_{+}^2}{4} K_{1}(\nu_{+} R) - 
    \frac{r_{1} \nu_{+}}{2} (K_{0}(\nu_{+} R) + K_{2}(\nu_{+} R)) \cos(\varphi_1)).\label{eq_series_K}
\end{split}
\end{align}
The first line in Eq.~(\ref{eq_series_K}) is just the effect of channel $1$. The
following lines represent the perturbation by the farfield around channel $2$. Inserting the expansions for $h$
into the boundary conditions and into the force and torque balance leads to lengthy systems
of linear equations for the coefficients of different order. The resulting first coefficients are approximately given by
\begin{align}
\begin{split}
&c^{k\pm}_0=\mp\frac{F_k + 2 b_k \alpha_k \kappab \pi \nu_{\pm}^2}{2 b_k \kappab \pi \nu_{\pm} (\nu_{+}^2 -\nu_{-}^2) K_1(b_k \nu_{\pm})},\\
&c^{1\pm}_1=\mp b_1^2 \nu_{\pm}\frac{(\nu_{-}^2 + 
   \nu_{+}^2) (\nu_{-} (F_2 + 2 b_2 \alpha_2 \kappab \pi \nu_{-}^2) K_1(R \nu_{-}) - 
   \nu_{+} (F_2 + 2 b_2 \alpha_2 \kappab \pi \nu_{+}^2) K_1(R \nu_{+}))}{4 \kappab \pi (\nu_{-}^2 - \nu_{+}^2)^2},\\
&c^{2\pm}_1=\pm b_2^2 \nu_{\pm}\frac{(\nu_{-}^2 + 
   \nu_{+}^2) (\nu_{-} (F_1 + 2 b_1 \alpha_1 \kappab \pi \nu_{-}^2) K_1(R \nu_{-}) -
   \nu_{+} (F_1 + 2 b_1 \alpha_1 \kappab \pi \nu_{+}^2) K_1(R \nu_{+}))}{4 \kappab \pi (\nu_{-}^2 - \nu_{+}^2)^2},
\end{split}
\end{align}
where terms up to the third power of the radii $(b_1^3,b_2^3,b_1^2b_2,b_2^2b_1)$ were taken into account. Errors in the given coefficients occur beyond this order.

To calculate the energy of the membrane, we employ Green's second identity and convert the surface integral in Eq.~(\ref{eq_Hs_surf}) to line integrals around the contours of the membrane
\begin{align}
\calH_{h,\epsilon} =\frac{\kappab}{2}\int\left[h\nabla^4 h-\xi^2 h\,\nabla h + \vartheta h^2\right]\,\rmd^2 r+
\frac{\kappab}{2}\oint (\mathbf{n} \cdot \nabla h)\nabla^2 h - h(\mathbf{n}\cdot\nabla\nabla^2h) + \xi^2 h (\mathbf{n}\cdot\nabla h) \rmd \mathbf{s},\label{eq_Hs}
\end{align}
where $\mathbf{n}$ is the unit normal vector on the contour pointing outwards. In equilibrium, Eq.~(\ref{eq_supp_mem}) holds and the surface integral in Eq.~(\ref{eq_Hs}) vanishes. Due to the assumption of an elastic membrane support, force acting on the channels is balanced by the whole system, not just at the outer perimeter of the membrane. Therefore, we can assume a flat membrane at the outer perimeter when $L\rightarrow \infty$. Thus, the energy becomes
\begin{align}
\begin{split}
\calH_{h,\epsilon}=\frac{\kappab}{2} \int\left[-\p_{r}h\,\nabla^2 h + h \p_r\nabla^2 h - \xi^2 h \p_r h \right] r \rmd \varphi_1 |_{r=b_1}+\text{contour at 2}.\label{eq_Hhhhh}
\end{split}
\end{align}
To evaluate the Laplace operators in Eq.~(\ref{eq_Hhhhh}) consistently, we employ $\nabla^2 h = \nu_{+}^2 h^{+} + \nu_{-}^2 h^{-}$. Leaving the energy $\calH_u$ of membrane thickness perturbations around individual channels aside, the membrane energy can be written as
\begin{align}
\calF_{m, \epsilon} =\calH_{h,\epsilon} +\gamma \pi (L^2 -b_1^2 -b_2^2) - F_1 h_1|_{r_1=0}- F_2 h_2|_{r_2=0}.\label{eq_Finterfull_epsilon}
\end{align}

We are not interested contributions of the energy that do not depend channel radii or conical angles. These are calculated from
Eq.~(\ref{eq_Finterfull_epsilon}) to be
\begin{equation}
\calF_{m, \epsilon}|_{\substack{b_{\{1,2\}}=0\\ \alpha_{\{1,2\}}=0}} = \gamma \pi L^2 + \frac{-2 F_1 F_2 K_0(R\nu_{-}) + 
  2 F_1 F_2 K_0(R \nu_+) + (F_1^2 + F_2^2) \log(\nu_-/\nu_+)}{
 4 \kappab \pi  (\nu_+^2-\nu_-^2)}.
\end{equation}
This energy is subtracted from the energy to obtain $\tcalF_{m, \epsilon} = \calF_{m, \epsilon}- \calF_{m, \epsilon}|_{\substack{b_{\{1,2\}}=0\\ \alpha_{\{1,2\}}=0}}$.

Next, we take the limit of vanishing elastic support $\epsilon \rightarrow 0$. From the definition~(\ref{eq_nudef}) of the lengthscales $\nu_{\pm}$ we then find $\nu^2_{+} \rightarrow \xi^2$ and $\nu^2_{-} \rightarrow 0$. For the energy we write
\begin{equation}
\tcalFm = \lim_{\epsilon \rightarrow 0}\tcalF_{m, \epsilon}.
\end{equation}
The full expression for $\tcalFm$ is lengthy, but can be somewhat abbreviated by subtracting the unperturbed energy of the single channels $\tcalF_{m,\{1,2\}}$ that can be calculated from Eq.~(\ref{eq_Overall_free_energy}). Keeping only terms up to cubic order in $b_k$ we find
\begin{align}
\begin{split}
\tcalF_{m,R} &= \tcalFm -\tcalF_{m,1}-\tcalF_{m, 2}\approx \\ 
&\, 2 \pi  b_1 \alpha_1 \alpha_2 b_2 \kappab \xi^2 K_0(R \xi)+K_0(R \xi) (b_1 \alpha_1
   F_2+\alpha_2 b_2 F_1)\\
&+\frac{1}{8} K_0(R \xi) \left(\xi^2 b_1^2 (2 X_1+1) +\xi^2 b_2^2
   (2 X_2+1)\right) (b_1 \alpha_1
   F_2+\alpha_2 b_2 F_1)\\
	&-\frac{b_1 b_2 \xi K_1(R \xi) (R \xi K_1(R
   \xi)-1) (b_1 \alpha_2 F_2+\alpha_1 b_2
   F_1)}{2 R}\\
	&-\frac{(R \xi K_1(R
   \xi)-1){}^2 \left(b_1^2 F_2^2+b_2^2 F_1^2\right)}{8
   \pi  \kappab R^2 \xi^2}
	+\frac{F_1 F_2 K_0(R \xi)
   \left(b_1^2 (2 X_1+1)+b_2^2 (2 X_2+1)\right)}{16 \pi 
   \kappab},\label{eq_full_interaction_enery}
\end{split}
\end{align}
where $X_k = -\Gamma_e - \log(\xi b_k/2)$. For $b_1=b_2$ and vanishing forces $F_{\{1,2\}}=0$, we recover the solution given in Ref.~\cite{weikl1998interaction} as
\begin{equation}
\tcalF_{m,R} \approx 2 \pi \kappab\, \alpha_1 \alpha_2 \, \xi^2 b_1^2 K_0(R \xi).
\end{equation}
When forces are present, we have up to linear order in channel radii
\begin{equation}
\tcalF_{m,R} \approx (\alpha_2 b_2 F_1 + b_1 \alpha_1 F_2) K_0(R\xi).
\end{equation}
The main text contains an expansion of Eq.~(\ref{eq_full_interaction_enery}) for the assumption $F_1=F_2$.
\begin{figure}[htb]
  \centering
	\includegraphics[scale=1.23]{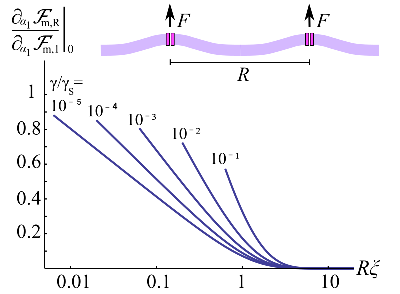} 
\caption{Two-channel interaction energy relative to the single-channel energy for infinitesimal conical deformation of channel 1. For close distances $R\xi \ll 1$, the interaction energy becomes comparable to the single-channel energy. Lines end at the minimum distance $R=(b_1+b_2)$. Parameters: $b_{\{1,2\}} =3\,\nm$, $\alpha_2=0$, $\kappab=25\,\kT$, $F\neq0$.} 
\label{fig_cooperativity}
\end{figure}

\begin{thebibliography}{66}%
\makeatletter
\providecommand \@ifxundefined [1]{%
 \@ifx{#1\undefined}
}%
\providecommand \@ifnum [1]{%
 \ifnum #1\expandafter \@firstoftwo
 \else \expandafter \@secondoftwo
 \fi
}%
\providecommand \@ifx [1]{%
 \ifx #1\expandafter \@firstoftwo
 \else \expandafter \@secondoftwo
 \fi
}%
\providecommand \natexlab [1]{#1}%
\providecommand \enquote  [1]{``#1''}%
\providecommand \bibnamefont  [1]{#1}%
\providecommand \bibfnamefont [1]{#1}%
\providecommand \citenamefont [1]{#1}%
\providecommand \href@noop [0]{\@secondoftwo}%
\providecommand \href [0]{\begingroup \@sanitize@url \@href}%
\providecommand \@href[1]{\@@startlink{#1}\@@href}%
\providecommand \@@href[1]{\endgroup#1\@@endlink}%
\providecommand \@sanitize@url [0]{\catcode `\\12\catcode `\$12\catcode
  `\&12\catcode `\#12\catcode `\^12\catcode `\_12\catcode `\%12\relax}%
\providecommand \@@startlink[1]{}%
\providecommand \@@endlink[0]{}%
\providecommand \url  [0]{\begingroup\@sanitize@url \@url }%
\providecommand \@url [1]{\endgroup\@href {#1}{\urlprefix }}%
\providecommand \urlprefix  [0]{URL }%
\providecommand \Eprint [0]{\href }%
\providecommand \doibase [0]{http://dx.doi.org/}%
\providecommand \selectlanguage [0]{\@gobble}%
\providecommand \bibinfo  [0]{\@secondoftwo}%
\providecommand \bibfield  [0]{\@secondoftwo}%
\providecommand \translation [1]{[#1]}%
\providecommand \BibitemOpen [0]{}%
\providecommand \bibitemStop [0]{}%
\providecommand \bibitemNoStop [0]{.\EOS\space}%
\providecommand \EOS [0]{\spacefactor3000\relax}%
\providecommand \BibitemShut  [1]{\csname bibitem#1\endcsname}%
\let\auto@bib@innerbib\@empty
\bibitem [{\citenamefont {Chalfie}(2009)}]{chalfie}%
  \BibitemOpen
  \bibfield  {author} {\bibinfo {author} {\bibfnamefont {M.}~\bibnamefont
  {Chalfie}},\ }\href@noop {} {\bibfield  {journal} {\bibinfo  {journal} {Nat.
  Rev. Mol. Cell Bio.}\ }\textbf {\bibinfo {volume} {10}},\ \bibinfo {pages}
  {44} (\bibinfo {year} {2009})}\BibitemShut {NoStop}%
\bibitem [{\citenamefont {Martinac}(2004)}]{martinac2004mechanosensitive}%
  \BibitemOpen
  \bibfield  {author} {\bibinfo {author} {\bibfnamefont {B.}~\bibnamefont
  {Martinac}},\ }\href@noop {} {\bibfield  {journal} {\bibinfo  {journal} {J.
  Cell Sci.}\ }\textbf {\bibinfo {volume} {117}},\ \bibinfo {pages} {2449}
  (\bibinfo {year} {2004})}\BibitemShut {NoStop}%
\bibitem [{\citenamefont {Hayakawa}\ \emph {et~al.}(2008)\citenamefont
  {Hayakawa}, \citenamefont {Tatsumi},\ and\ \citenamefont
  {Sokabe}}]{hayakawa2008actin}%
  \BibitemOpen
  \bibfield  {author} {\bibinfo {author} {\bibfnamefont {K.}~\bibnamefont
  {Hayakawa}}, \bibinfo {author} {\bibfnamefont {H.}~\bibnamefont {Tatsumi}}, \
  and\ \bibinfo {author} {\bibfnamefont {M.}~\bibnamefont {Sokabe}},\
  }\href@noop {} {\bibfield  {journal} {\bibinfo  {journal} {J. Cell Sci.}\
  }\textbf {\bibinfo {volume} {121}},\ \bibinfo {pages} {496} (\bibinfo {year}
  {2008})}\BibitemShut {NoStop}%
\bibitem [{\citenamefont {Bass}\ \emph {et~al.}(2002)\citenamefont {Bass},
  \citenamefont {Strop}, \citenamefont {Barclay},\ and\ \citenamefont
  {Rees}}]{bass2002crystal}%
  \BibitemOpen
  \bibfield  {author} {\bibinfo {author} {\bibfnamefont {R.~B.}\ \bibnamefont
  {Bass}}, \bibinfo {author} {\bibfnamefont {P.}~\bibnamefont {Strop}},
  \bibinfo {author} {\bibfnamefont {M.}~\bibnamefont {Barclay}}, \ and\
  \bibinfo {author} {\bibfnamefont {D.~C.}\ \bibnamefont {Rees}},\ }\href@noop
  {} {\bibfield  {journal} {\bibinfo  {journal} {Science}\ }\textbf {\bibinfo
  {volume} {298}},\ \bibinfo {pages} {1582} (\bibinfo {year}
  {2002})}\BibitemShut {NoStop}%
\bibitem [{\citenamefont {Sukharev}\ and\ \citenamefont
  {Corey}(2004)}]{sukharev2004mechanosensitive}%
  \BibitemOpen
  \bibfield  {author} {\bibinfo {author} {\bibfnamefont {S.}~\bibnamefont
  {Sukharev}}\ and\ \bibinfo {author} {\bibfnamefont {D.~P.}\ \bibnamefont
  {Corey}},\ }\href@noop {} {\bibfield  {journal} {\bibinfo  {journal} {Sci.
  Signal.}\ }\textbf {\bibinfo {volume} {2004}},\ \bibinfo {pages} {re4}
  (\bibinfo {year} {2004})}\BibitemShut {NoStop}%
\bibitem [{\citenamefont {Phillips}\ \emph {et~al.}(2009)\citenamefont
  {Phillips}, \citenamefont {Ursell}, \citenamefont {Wiggins},\ and\
  \citenamefont {Sens}}]{phillips2009emerging}%
  \BibitemOpen
  \bibfield  {author} {\bibinfo {author} {\bibfnamefont {R.}~\bibnamefont
  {Phillips}}, \bibinfo {author} {\bibfnamefont {T.}~\bibnamefont {Ursell}},
  \bibinfo {author} {\bibfnamefont {P.}~\bibnamefont {Wiggins}}, \ and\
  \bibinfo {author} {\bibfnamefont {P.}~\bibnamefont {Sens}},\ }\href@noop {}
  {\bibfield  {journal} {\bibinfo  {journal} {Nature}\ }\textbf {\bibinfo
  {volume} {459}},\ \bibinfo {pages} {379} (\bibinfo {year}
  {2009})}\BibitemShut {NoStop}%
\bibitem [{\citenamefont {Dan}\ and\ \citenamefont
  {Safran}(1998)}]{dan1998effect}%
  \BibitemOpen
  \bibfield  {author} {\bibinfo {author} {\bibfnamefont {N.}~\bibnamefont
  {Dan}}\ and\ \bibinfo {author} {\bibfnamefont {S.~A.}\ \bibnamefont
  {Safran}},\ }\href@noop {} {\bibfield  {journal} {\bibinfo  {journal}
  {Biophys. J.}\ }\textbf {\bibinfo {volume} {75}},\ \bibinfo {pages} {1410}
  (\bibinfo {year} {1998})}\BibitemShut {NoStop}%
\bibitem [{\citenamefont {Turner}\ and\ \citenamefont
  {Sens}(2004)}]{turner2004gating}%
  \BibitemOpen
  \bibfield  {author} {\bibinfo {author} {\bibfnamefont {M.~S.}\ \bibnamefont
  {Turner}}\ and\ \bibinfo {author} {\bibfnamefont {P.}~\bibnamefont {Sens}},\
  }\href@noop {} {\bibfield  {journal} {\bibinfo  {journal} {Phys. Rev. Lett.}\
  }\textbf {\bibinfo {volume} {93}},\ \bibinfo {pages} {118103} (\bibinfo
  {year} {2004})}\BibitemShut {NoStop}%
\bibitem [{\citenamefont {Lee}(2006)}]{lee2006energetics}%
  \BibitemOpen
  \bibfield  {author} {\bibinfo {author} {\bibfnamefont {K.-J.-B.}\
  \bibnamefont {Lee}},\ }\href@noop {} {\bibfield  {journal} {\bibinfo
  {journal} {Phys. Rev. E}\ }\textbf {\bibinfo {volume} {73}},\ \bibinfo
  {pages} {021909} (\bibinfo {year} {2006})}\BibitemShut {NoStop}%
\bibitem [{\citenamefont {Chen}\ \emph {et~al.}(2008)\citenamefont {Chen},
  \citenamefont {Cui}, \citenamefont {Tang}, \citenamefont {Yoo},\ and\
  \citenamefont {Yethiraj}}]{chen2008gating}%
  \BibitemOpen
  \bibfield  {author} {\bibinfo {author} {\bibfnamefont {X.}~\bibnamefont
  {Chen}}, \bibinfo {author} {\bibfnamefont {Q.}~\bibnamefont {Cui}}, \bibinfo
  {author} {\bibfnamefont {Y.}~\bibnamefont {Tang}}, \bibinfo {author}
  {\bibfnamefont {J.}~\bibnamefont {Yoo}}, \ and\ \bibinfo {author}
  {\bibfnamefont {A.}~\bibnamefont {Yethiraj}},\ }\href@noop {} {\bibfield
  {journal} {\bibinfo  {journal} {Biophys. J.}\ }\textbf {\bibinfo {volume}
  {95}},\ \bibinfo {pages} {563} (\bibinfo {year} {2008})}\BibitemShut
  {NoStop}%
\bibitem [{\citenamefont {Rautu}\ \emph {et~al.}(2015)\citenamefont {Rautu},
  \citenamefont {Rowlands},\ and\ \citenamefont {Turner}}]{rautu2015membrane}%
  \BibitemOpen
  \bibfield  {author} {\bibinfo {author} {\bibfnamefont {S.~A.}\ \bibnamefont
  {Rautu}}, \bibinfo {author} {\bibfnamefont {G.}~\bibnamefont {Rowlands}}, \
  and\ \bibinfo {author} {\bibfnamefont {M.~S.}\ \bibnamefont {Turner}},\
  }\href@noop {} {\bibfield  {journal} {\bibinfo  {journal} {Phys. Rev. Lett.}\
  }\textbf {\bibinfo {volume} {114}},\ \bibinfo {pages} {098101} (\bibinfo
  {year} {2015})}\BibitemShut {NoStop}%
\bibitem [{\citenamefont {Wiggins}\ and\ \citenamefont
  {Phillips}(2004)}]{wiggins2004analytic}%
  \BibitemOpen
  \bibfield  {author} {\bibinfo {author} {\bibfnamefont {P.}~\bibnamefont
  {Wiggins}}\ and\ \bibinfo {author} {\bibfnamefont {R.}~\bibnamefont
  {Phillips}},\ }\href@noop {} {\bibfield  {journal} {\bibinfo  {journal}
  {Proc. Nat. Acad. Sci.}\ }\textbf {\bibinfo {volume} {101}},\ \bibinfo
  {pages} {4071} (\bibinfo {year} {2004})}\BibitemShut {NoStop}%
\bibitem [{\citenamefont {Markin}\ and\ \citenamefont
  {Sachs}(2004)}]{markin2004thermodynamics}%
  \BibitemOpen
  \bibfield  {author} {\bibinfo {author} {\bibfnamefont {V.}~\bibnamefont
  {Markin}}\ and\ \bibinfo {author} {\bibfnamefont {F.}~\bibnamefont {Sachs}},\
  }\href@noop {} {\bibfield  {journal} {\bibinfo  {journal} {Phys. Biol.}\
  }\textbf {\bibinfo {volume} {1}},\ \bibinfo {pages} {110} (\bibinfo {year}
  {2004})}\BibitemShut {NoStop}%
\bibitem [{\citenamefont {Reeves}\ \emph {et~al.}(2008)\citenamefont {Reeves},
  \citenamefont {Ursell}, \citenamefont {Sens}, \citenamefont {Kondev},\ and\
  \citenamefont {Phillips}}]{reeves2008membrane}%
  \BibitemOpen
  \bibfield  {author} {\bibinfo {author} {\bibfnamefont {D.}~\bibnamefont
  {Reeves}}, \bibinfo {author} {\bibfnamefont {T.}~\bibnamefont {Ursell}},
  \bibinfo {author} {\bibfnamefont {P.}~\bibnamefont {Sens}}, \bibinfo {author}
  {\bibfnamefont {J.}~\bibnamefont {Kondev}}, \ and\ \bibinfo {author}
  {\bibfnamefont {R.}~\bibnamefont {Phillips}},\ }\href@noop {} {\bibfield
  {journal} {\bibinfo  {journal} {Phys. Rev. E}\ }\textbf {\bibinfo {volume}
  {78}},\ \bibinfo {pages} {041901} (\bibinfo {year} {2008})}\BibitemShut
  {NoStop}%
\bibitem [{\citenamefont {Pak}\ \emph {et~al.}(2015)\citenamefont {Pak},
  \citenamefont {Young}, \citenamefont {Marple}, \citenamefont {Veerapaneni},\
  and\ \citenamefont {Stone}}]{pak2015gating}%
  \BibitemOpen
  \bibfield  {author} {\bibinfo {author} {\bibfnamefont {O.~S.}\ \bibnamefont
  {Pak}}, \bibinfo {author} {\bibfnamefont {Y.-N.}\ \bibnamefont {Young}},
  \bibinfo {author} {\bibfnamefont {G.~R.}\ \bibnamefont {Marple}}, \bibinfo
  {author} {\bibfnamefont {S.}~\bibnamefont {Veerapaneni}}, \ and\ \bibinfo
  {author} {\bibfnamefont {H.~A.}\ \bibnamefont {Stone}},\ }\href@noop {}
  {\bibfield  {journal} {\bibinfo  {journal} {Proc. Nat. Acad. Sci.}\ }\textbf
  {\bibinfo {volume} {112}},\ \bibinfo {pages} {9822} (\bibinfo {year}
  {2015})}\BibitemShut {NoStop}%
\bibitem [{\citenamefont {Huang}(1986)}]{huang1986deformation}%
  \BibitemOpen
  \bibfield  {author} {\bibinfo {author} {\bibfnamefont {H.~W.}\ \bibnamefont
  {Huang}},\ }\href@noop {} {\bibfield  {journal} {\bibinfo  {journal}
  {Biophys. J.}\ }\textbf {\bibinfo {volume} {50}},\ \bibinfo {pages} {1061}
  (\bibinfo {year} {1986})}\BibitemShut {NoStop}%
\bibitem [{\citenamefont {Helfrich}\ and\ \citenamefont
  {Jakobsson}(1990)}]{helfrich1990calculation}%
  \BibitemOpen
  \bibfield  {author} {\bibinfo {author} {\bibfnamefont {P.}~\bibnamefont
  {Helfrich}}\ and\ \bibinfo {author} {\bibfnamefont {E.}~\bibnamefont
  {Jakobsson}},\ }\href@noop {} {\bibfield  {journal} {\bibinfo  {journal}
  {Biophys. J.}\ }\textbf {\bibinfo {volume} {57}},\ \bibinfo {pages} {1075}
  (\bibinfo {year} {1990})}\BibitemShut {NoStop}%
\bibitem [{\citenamefont {Kung}(2005)}]{kung2005possible}%
  \BibitemOpen
  \bibfield  {author} {\bibinfo {author} {\bibfnamefont {C.}~\bibnamefont
  {Kung}},\ }\href@noop {} {\bibfield  {journal} {\bibinfo  {journal} {Nature}\
  }\textbf {\bibinfo {volume} {436}},\ \bibinfo {pages} {647} (\bibinfo {year}
  {2005})}\BibitemShut {NoStop}%
\bibitem [{\citenamefont {Orr}\ \emph {et~al.}(2006)\citenamefont {Orr},
  \citenamefont {Helmke}, \citenamefont {Blackman},\ and\ \citenamefont
  {Schwartz}}]{orr2006mechanisms}%
  \BibitemOpen
  \bibfield  {author} {\bibinfo {author} {\bibfnamefont {A.~W.}\ \bibnamefont
  {Orr}}, \bibinfo {author} {\bibfnamefont {B.~P.}\ \bibnamefont {Helmke}},
  \bibinfo {author} {\bibfnamefont {B.~R.}\ \bibnamefont {Blackman}}, \ and\
  \bibinfo {author} {\bibfnamefont {M.~A.}\ \bibnamefont {Schwartz}},\
  }\href@noop {} {\bibfield  {journal} {\bibinfo  {journal} {Dev. Cell}\
  }\textbf {\bibinfo {volume} {10}},\ \bibinfo {pages} {11} (\bibinfo {year}
  {2006})}\BibitemShut {NoStop}%
\bibitem [{\citenamefont {Martinac}(2014)}]{martinac2014ion}%
  \BibitemOpen
  \bibfield  {author} {\bibinfo {author} {\bibfnamefont {B.}~\bibnamefont
  {Martinac}},\ }\href@noop {} {\bibfield  {journal} {\bibinfo  {journal}
  {Biochim. Biophys. Acta}\ }\textbf {\bibinfo {volume} {1838}},\ \bibinfo
  {pages} {682} (\bibinfo {year} {2014})}\BibitemShut {NoStop}%
\bibitem [{\citenamefont {Tavernarakis}\ and\ \citenamefont
  {Driscoll}(1997)}]{tavernarakis1997molecular}%
  \BibitemOpen
  \bibfield  {author} {\bibinfo {author} {\bibfnamefont {N.}~\bibnamefont
  {Tavernarakis}}\ and\ \bibinfo {author} {\bibfnamefont {M.}~\bibnamefont
  {Driscoll}},\ }\href@noop {} {\bibfield  {journal} {\bibinfo  {journal}
  {Annu. Rev. Physiol.}\ }\textbf {\bibinfo {volume} {59}},\ \bibinfo {pages}
  {659} (\bibinfo {year} {1997})}\BibitemShut {NoStop}%
\bibitem [{\citenamefont {O'Hagan}\ \emph {et~al.}(2005)\citenamefont
  {O'Hagan}, \citenamefont {Chalfie},\ and\ \citenamefont
  {Goodman}}]{o2005mec}%
  \BibitemOpen
  \bibfield  {author} {\bibinfo {author} {\bibfnamefont {R.}~\bibnamefont
  {O'Hagan}}, \bibinfo {author} {\bibfnamefont {M.}~\bibnamefont {Chalfie}}, \
  and\ \bibinfo {author} {\bibfnamefont {M.~B.}\ \bibnamefont {Goodman}},\
  }\href@noop {} {\bibfield  {journal} {\bibinfo  {journal} {Nat. Neurosci.}\
  }\textbf {\bibinfo {volume} {8}},\ \bibinfo {pages} {43} (\bibinfo {year}
  {2005})}\BibitemShut {NoStop}%
\bibitem [{\citenamefont {Howard}\ and\ \citenamefont
  {Bechstedt}(2004)}]{howard2004hypothesis}%
  \BibitemOpen
  \bibfield  {author} {\bibinfo {author} {\bibfnamefont {J.}~\bibnamefont
  {Howard}}\ and\ \bibinfo {author} {\bibfnamefont {S.}~\bibnamefont
  {Bechstedt}},\ }\href@noop {} {\bibfield  {journal} {\bibinfo  {journal}
  {Curr. Biol.}\ }\textbf {\bibinfo {volume} {14}},\ \bibinfo {pages} {R224}
  (\bibinfo {year} {2004})}\BibitemShut {NoStop}%
\bibitem [{\citenamefont {Delmas}\ and\ \citenamefont
  {Coste}(2013)}]{delmas2013mechano}%
  \BibitemOpen
  \bibfield  {author} {\bibinfo {author} {\bibfnamefont {P.}~\bibnamefont
  {Delmas}}\ and\ \bibinfo {author} {\bibfnamefont {B.}~\bibnamefont {Coste}},\
  }\href@noop {} {\bibfield  {journal} {\bibinfo  {journal} {Cell}\ }\textbf
  {\bibinfo {volume} {155}},\ \bibinfo {pages} {278} (\bibinfo {year}
  {2013})}\BibitemShut {NoStop}%
\bibitem [{\citenamefont {Liu}\ and\ \citenamefont
  {Montell}(2015)}]{liu2015forcing}%
  \BibitemOpen
  \bibfield  {author} {\bibinfo {author} {\bibfnamefont {C.}~\bibnamefont
  {Liu}}\ and\ \bibinfo {author} {\bibfnamefont {C.}~\bibnamefont {Montell}},\
  }\href@noop {} {\bibfield  {journal} {\bibinfo  {journal} {Biochem. Biophys.
  Res. Com.}\ }\textbf {\bibinfo {volume} {460}},\ \bibinfo {pages} {22}
  (\bibinfo {year} {2015})}\BibitemShut {NoStop}%
\bibitem [{\citenamefont {Zhang}\ \emph {et~al.}(2015)\citenamefont {Zhang},
  \citenamefont {Cheng}, \citenamefont {Kittelmann}, \citenamefont {Li},
  \citenamefont {Petkovic}, \citenamefont {Cheng}, \citenamefont {Jin},
  \citenamefont {Guo}, \citenamefont {G{\"o}pfert}, \citenamefont {Jan} \emph
  {et~al.}}]{zhang2015ankyrin}%
  \BibitemOpen
  \bibfield  {author} {\bibinfo {author} {\bibfnamefont {W.}~\bibnamefont
  {Zhang}}, \bibinfo {author} {\bibfnamefont {L.~E.}\ \bibnamefont {Cheng}},
  \bibinfo {author} {\bibfnamefont {M.}~\bibnamefont {Kittelmann}}, \bibinfo
  {author} {\bibfnamefont {J.}~\bibnamefont {Li}}, \bibinfo {author}
  {\bibfnamefont {M.}~\bibnamefont {Petkovic}}, \bibinfo {author}
  {\bibfnamefont {T.}~\bibnamefont {Cheng}}, \bibinfo {author} {\bibfnamefont
  {P.}~\bibnamefont {Jin}}, \bibinfo {author} {\bibfnamefont {Z.}~\bibnamefont
  {Guo}}, \bibinfo {author} {\bibfnamefont {M.~C.}\ \bibnamefont
  {G{\"o}pfert}}, \bibinfo {author} {\bibfnamefont {L.~Y.}\ \bibnamefont
  {Jan}},  \emph {et~al.},\ }\href@noop {} {\bibfield  {journal} {\bibinfo
  {journal} {Cell}\ }\textbf {\bibinfo {volume} {162}},\ \bibinfo {pages}
  {1391} (\bibinfo {year} {2015})}\BibitemShut {NoStop}%
\bibitem [{\citenamefont {Sotomayor}\ \emph {et~al.}(2005)\citenamefont
  {Sotomayor}, \citenamefont {Corey},\ and\ \citenamefont
  {Schulten}}]{sotomayor2005search}%
  \BibitemOpen
  \bibfield  {author} {\bibinfo {author} {\bibfnamefont {M.}~\bibnamefont
  {Sotomayor}}, \bibinfo {author} {\bibfnamefont {D.~P.}\ \bibnamefont
  {Corey}}, \ and\ \bibinfo {author} {\bibfnamefont {K.}~\bibnamefont
  {Schulten}},\ }\href@noop {} {\bibfield  {journal} {\bibinfo  {journal}
  {Structure}\ }\textbf {\bibinfo {volume} {13}},\ \bibinfo {pages} {669}
  (\bibinfo {year} {2005})}\BibitemShut {NoStop}%
\bibitem [{\citenamefont {Li}\ \emph {et~al.}(2006)\citenamefont {Li},
  \citenamefont {Wetzel}, \citenamefont {Pl{\"u}ckthun},\ and\ \citenamefont
  {Fernandez}}]{li2006stepwise}%
  \BibitemOpen
  \bibfield  {author} {\bibinfo {author} {\bibfnamefont {L.}~\bibnamefont
  {Li}}, \bibinfo {author} {\bibfnamefont {S.}~\bibnamefont {Wetzel}}, \bibinfo
  {author} {\bibfnamefont {A.}~\bibnamefont {Pl{\"u}ckthun}}, \ and\ \bibinfo
  {author} {\bibfnamefont {J.~M.}\ \bibnamefont {Fernandez}},\ }\href@noop {}
  {\bibfield  {journal} {\bibinfo  {journal} {Biophys. J.}\ }\textbf {\bibinfo
  {volume} {90}},\ \bibinfo {pages} {L30} (\bibinfo {year} {2006})}\BibitemShut
  {NoStop}%
\bibitem [{\citenamefont {Prager-Khoutorsky}\ \emph {et~al.}(2014)\citenamefont
  {Prager-Khoutorsky}, \citenamefont {Khoutorsky},\ and\ \citenamefont
  {Bourque}}]{prager2014unique}%
  \BibitemOpen
  \bibfield  {author} {\bibinfo {author} {\bibfnamefont {M.}~\bibnamefont
  {Prager-Khoutorsky}}, \bibinfo {author} {\bibfnamefont {A.}~\bibnamefont
  {Khoutorsky}}, \ and\ \bibinfo {author} {\bibfnamefont {C.~W.}\ \bibnamefont
  {Bourque}},\ }\href@noop {} {\bibfield  {journal} {\bibinfo  {journal}
  {Neuron}\ }\textbf {\bibinfo {volume} {83}},\ \bibinfo {pages} {866}
  (\bibinfo {year} {2014})}\BibitemShut {NoStop}%
\bibitem [{\citenamefont {Hill}\ and\ \citenamefont
  {Schaefer}(2007)}]{hill2007trpa1}%
  \BibitemOpen
  \bibfield  {author} {\bibinfo {author} {\bibfnamefont {K.}~\bibnamefont
  {Hill}}\ and\ \bibinfo {author} {\bibfnamefont {M.}~\bibnamefont
  {Schaefer}},\ }\href@noop {} {\bibfield  {journal} {\bibinfo  {journal} {J.
  Biol. Chem.}\ }\textbf {\bibinfo {volume} {282}},\ \bibinfo {pages} {7145}
  (\bibinfo {year} {2007})}\BibitemShut {NoStop}%
\bibitem [{\citenamefont {Suchyna}\ \emph {et~al.}(2004)\citenamefont
  {Suchyna}, \citenamefont {Tape}, \citenamefont {Koeppe}, \citenamefont
  {Andersen}, \citenamefont {Sachs},\ and\ \citenamefont
  {Gottlieb}}]{suchyna2004bilayer}%
  \BibitemOpen
  \bibfield  {author} {\bibinfo {author} {\bibfnamefont {T.~M.}\ \bibnamefont
  {Suchyna}}, \bibinfo {author} {\bibfnamefont {S.~E.}\ \bibnamefont {Tape}},
  \bibinfo {author} {\bibfnamefont {R.~E.}\ \bibnamefont {Koeppe}}, \bibinfo
  {author} {\bibfnamefont {O.~S.}\ \bibnamefont {Andersen}}, \bibinfo {author}
  {\bibfnamefont {F.}~\bibnamefont {Sachs}}, \ and\ \bibinfo {author}
  {\bibfnamefont {P.~A.}\ \bibnamefont {Gottlieb}},\ }\href@noop {} {\bibfield
  {journal} {\bibinfo  {journal} {Nature}\ }\textbf {\bibinfo {volume} {430}},\
  \bibinfo {pages} {235} (\bibinfo {year} {2004})}\BibitemShut {NoStop}%
\bibitem [{\citenamefont {Khalili-Araghi}\ \emph {et~al.}(2009)\citenamefont
  {Khalili-Araghi}, \citenamefont {Gumbart}, \citenamefont {Wen}, \citenamefont
  {Sotomayor}, \citenamefont {Tajkhorshid},\ and\ \citenamefont
  {Schulten}}]{khalili2009molecular}%
  \BibitemOpen
  \bibfield  {author} {\bibinfo {author} {\bibfnamefont {F.}~\bibnamefont
  {Khalili-Araghi}}, \bibinfo {author} {\bibfnamefont {J.}~\bibnamefont
  {Gumbart}}, \bibinfo {author} {\bibfnamefont {P.-C.}\ \bibnamefont {Wen}},
  \bibinfo {author} {\bibfnamefont {M.}~\bibnamefont {Sotomayor}}, \bibinfo
  {author} {\bibfnamefont {E.}~\bibnamefont {Tajkhorshid}}, \ and\ \bibinfo
  {author} {\bibfnamefont {K.}~\bibnamefont {Schulten}},\ }\href@noop {}
  {\bibfield  {journal} {\bibinfo  {journal} {Curr. Opin. Struct. Biol.}\
  }\textbf {\bibinfo {volume} {19}},\ \bibinfo {pages} {128} (\bibinfo {year}
  {2009})}\BibitemShut {NoStop}%
\bibitem [{\citenamefont {Rawicz}\ \emph {et~al.}(2000)\citenamefont {Rawicz},
  \citenamefont {Olbrich}, \citenamefont {McIntosh}, \citenamefont {Needham},\
  and\ \citenamefont {Evans}}]{rawicz2000effect}%
  \BibitemOpen
  \bibfield  {author} {\bibinfo {author} {\bibfnamefont {W.}~\bibnamefont
  {Rawicz}}, \bibinfo {author} {\bibfnamefont {K.}~\bibnamefont {Olbrich}},
  \bibinfo {author} {\bibfnamefont {T.}~\bibnamefont {McIntosh}}, \bibinfo
  {author} {\bibfnamefont {D.}~\bibnamefont {Needham}}, \ and\ \bibinfo
  {author} {\bibfnamefont {E.}~\bibnamefont {Evans}},\ }\href@noop {}
  {\bibfield  {journal} {\bibinfo  {journal} {Biophys. J.}\ }\textbf {\bibinfo
  {volume} {79}},\ \bibinfo {pages} {328} (\bibinfo {year} {2000})}\BibitemShut
  {NoStop}%
\bibitem [{Sup()}]{Suppl_mat}%
  \BibitemOpen
  \href@noop {} {}\bibinfo {note} {See supplemental material.}\BibitemShut
  {Stop}%
\bibitem [{\citenamefont {Dai}\ and\ \citenamefont
  {Sheetz}(1999)}]{dai1999membrane}%
  \BibitemOpen
  \bibfield  {author} {\bibinfo {author} {\bibfnamefont {J.}~\bibnamefont
  {Dai}}\ and\ \bibinfo {author} {\bibfnamefont {M.~P.}\ \bibnamefont
  {Sheetz}},\ }\href@noop {} {\bibfield  {journal} {\bibinfo  {journal}
  {Biophys. J.}\ }\textbf {\bibinfo {volume} {77}},\ \bibinfo {pages} {3363}
  (\bibinfo {year} {1999})}\BibitemShut {NoStop}%
\bibitem [{\citenamefont {Raucher}\ and\ \citenamefont
  {Sheetz}(1999)}]{raucher1999characteristics}%
  \BibitemOpen
  \bibfield  {author} {\bibinfo {author} {\bibfnamefont {D.}~\bibnamefont
  {Raucher}}\ and\ \bibinfo {author} {\bibfnamefont {M.~P.}\ \bibnamefont
  {Sheetz}},\ }\href@noop {} {\bibfield  {journal} {\bibinfo  {journal}
  {Biophys. J.}\ }\textbf {\bibinfo {volume} {77}},\ \bibinfo {pages} {1992}
  (\bibinfo {year} {1999})}\BibitemShut {NoStop}%
\bibitem [{\citenamefont {Der{\'e}nyi}\ \emph {et~al.}(2002)\citenamefont
  {Der{\'e}nyi}, \citenamefont {J{\"u}licher},\ and\ \citenamefont
  {Prost}}]{derenyi2002formation}%
  \BibitemOpen
  \bibfield  {author} {\bibinfo {author} {\bibfnamefont {I.}~\bibnamefont
  {Der{\'e}nyi}}, \bibinfo {author} {\bibfnamefont {F.}~\bibnamefont
  {J{\"u}licher}}, \ and\ \bibinfo {author} {\bibfnamefont {J.}~\bibnamefont
  {Prost}},\ }\href@noop {} {\bibfield  {journal} {\bibinfo  {journal} {Phys.
  Rev. Lett.}\ }\textbf {\bibinfo {volume} {88}},\ \bibinfo {pages} {238101}
  (\bibinfo {year} {2002})}\BibitemShut {NoStop}%
\bibitem [{\citenamefont {Jablin}\ \emph {et~al.}(2014)\citenamefont {Jablin},
  \citenamefont {Akabori},\ and\ \citenamefont
  {Nagle}}]{jablin2014experimental}%
  \BibitemOpen
  \bibfield  {author} {\bibinfo {author} {\bibfnamefont {M.}~\bibnamefont
  {Jablin}}, \bibinfo {author} {\bibfnamefont {K.}~\bibnamefont {Akabori}}, \
  and\ \bibinfo {author} {\bibfnamefont {J.}~\bibnamefont {Nagle}},\
  }\href@noop {} {\bibfield  {journal} {\bibinfo  {journal} {Phys. Rev. Lett.}\
  }\textbf {\bibinfo {volume} {113}},\ \bibinfo {pages} {248102} (\bibinfo
  {year} {2014})}\BibitemShut {NoStop}%
\bibitem [{\citenamefont {Jensen}\ and\ \citenamefont
  {Mouritsen}(2004)}]{jensen2004lipids}%
  \BibitemOpen
  \bibfield  {author} {\bibinfo {author} {\bibfnamefont {M.~{\O}.}\
  \bibnamefont {Jensen}}\ and\ \bibinfo {author} {\bibfnamefont {O.~G.}\
  \bibnamefont {Mouritsen}},\ }\href@noop {} {\bibfield  {journal} {\bibinfo
  {journal} {Biochim. Biophys. Acta}\ }\textbf {\bibinfo {volume} {1666}},\
  \bibinfo {pages} {205} (\bibinfo {year} {2004})}\BibitemShut {NoStop}%
\bibitem [{\citenamefont {Andersen}\ and\ \citenamefont
  {Koeppe}(2007)}]{andersen2007bilayer}%
  \BibitemOpen
  \bibfield  {author} {\bibinfo {author} {\bibfnamefont {O.~S.}\ \bibnamefont
  {Andersen}}\ and\ \bibinfo {author} {\bibfnamefont {R.~E.}\ \bibnamefont
  {Koeppe}},\ }\href@noop {} {\bibfield  {journal} {\bibinfo  {journal} {Annu.
  Rev. Biophys. Biomol. Struct.}\ }\textbf {\bibinfo {volume} {36}},\ \bibinfo
  {pages} {107} (\bibinfo {year} {2007})}\BibitemShut {NoStop}%
\bibitem [{\citenamefont {MacKintosh}\ and\ \citenamefont
  {Lubensky}(1991)}]{mackintosh1991orientational}%
  \BibitemOpen
  \bibfield  {author} {\bibinfo {author} {\bibfnamefont {F.}~\bibnamefont
  {MacKintosh}}\ and\ \bibinfo {author} {\bibfnamefont {T.}~\bibnamefont
  {Lubensky}},\ }\href@noop {} {\bibfield  {journal} {\bibinfo  {journal}
  {Phys. Rev. Lett.}\ }\textbf {\bibinfo {volume} {67}},\ \bibinfo {pages}
  {1169} (\bibinfo {year} {1991})}\BibitemShut {NoStop}%
\bibitem [{\citenamefont {Seifert}\ \emph {et~al.}(1996)\citenamefont
  {Seifert}, \citenamefont {Shillcock},\ and\ \citenamefont
  {Nelson}}]{seifert1996role}%
  \BibitemOpen
  \bibfield  {author} {\bibinfo {author} {\bibfnamefont {U.}~\bibnamefont
  {Seifert}}, \bibinfo {author} {\bibfnamefont {J.}~\bibnamefont {Shillcock}},
  \ and\ \bibinfo {author} {\bibfnamefont {P.}~\bibnamefont {Nelson}},\
  }\href@noop {} {\bibfield  {journal} {\bibinfo  {journal} {Phys. Rev. Lett.}\
  }\textbf {\bibinfo {volume} {77}},\ \bibinfo {pages} {5237} (\bibinfo {year}
  {1996})}\BibitemShut {NoStop}%
\bibitem [{\citenamefont {Fournier}(1999)}]{fournier1999microscopic}%
  \BibitemOpen
  \bibfield  {author} {\bibinfo {author} {\bibfnamefont {J.-B.}\ \bibnamefont
  {Fournier}},\ }\href@noop {} {\bibfield  {journal} {\bibinfo  {journal} {Eur.
  Phys. J. E}\ }\textbf {\bibinfo {volume} {11}},\ \bibinfo {pages} {261}
  (\bibinfo {year} {1999})}\BibitemShut {NoStop}%
\bibitem [{\citenamefont {Kozlovsky}\ \emph {et~al.}(2004)\citenamefont
  {Kozlovsky}, \citenamefont {Zimmerberg},\ and\ \citenamefont
  {Kozlov}}]{kozlovsky2004orientation}%
  \BibitemOpen
  \bibfield  {author} {\bibinfo {author} {\bibfnamefont {Y.}~\bibnamefont
  {Kozlovsky}}, \bibinfo {author} {\bibfnamefont {J.}~\bibnamefont
  {Zimmerberg}}, \ and\ \bibinfo {author} {\bibfnamefont {M.~M.}\ \bibnamefont
  {Kozlov}},\ }\href@noop {} {\bibfield  {journal} {\bibinfo  {journal}
  {Biophys. J.}\ }\textbf {\bibinfo {volume} {87}},\ \bibinfo {pages} {999}
  (\bibinfo {year} {2004})}\BibitemShut {NoStop}%
\bibitem [{\citenamefont {May}\ \emph {et~al.}(2004)\citenamefont {May},
  \citenamefont {Kozlovsky}, \citenamefont {Ben-Shaul},\ and\ \citenamefont
  {Kozlov}}]{may2004tilt}%
  \BibitemOpen
  \bibfield  {author} {\bibinfo {author} {\bibfnamefont {S.}~\bibnamefont
  {May}}, \bibinfo {author} {\bibfnamefont {Y.}~\bibnamefont {Kozlovsky}},
  \bibinfo {author} {\bibfnamefont {A.}~\bibnamefont {Ben-Shaul}}, \ and\
  \bibinfo {author} {\bibfnamefont {M.}~\bibnamefont {Kozlov}},\ }\href@noop {}
  {\bibfield  {journal} {\bibinfo  {journal} {Eur. Phys. J. E}\ }\textbf
  {\bibinfo {volume} {14}},\ \bibinfo {pages} {299} (\bibinfo {year}
  {2004})}\BibitemShut {NoStop}%
\bibitem [{\citenamefont {Kuzmin}\ \emph {et~al.}(2005)\citenamefont {Kuzmin},
  \citenamefont {Akimov}, \citenamefont {Chizmadzhev}, \citenamefont
  {Zimmerberg},\ and\ \citenamefont {Cohen}}]{kuzmin2005line}%
  \BibitemOpen
  \bibfield  {author} {\bibinfo {author} {\bibfnamefont {P.~I.}\ \bibnamefont
  {Kuzmin}}, \bibinfo {author} {\bibfnamefont {S.~A.}\ \bibnamefont {Akimov}},
  \bibinfo {author} {\bibfnamefont {Y.~A.}\ \bibnamefont {Chizmadzhev}},
  \bibinfo {author} {\bibfnamefont {J.}~\bibnamefont {Zimmerberg}}, \ and\
  \bibinfo {author} {\bibfnamefont {F.~S.}\ \bibnamefont {Cohen}},\ }\href@noop
  {} {\bibfield  {journal} {\bibinfo  {journal} {Biophys. J.}\ }\textbf
  {\bibinfo {volume} {88}},\ \bibinfo {pages} {1120} (\bibinfo {year}
  {2005})}\BibitemShut {NoStop}%
\bibitem [{\citenamefont {Venturoli}\ \emph {et~al.}(2006)\citenamefont
  {Venturoli}, \citenamefont {Sperotto}, \citenamefont {Kranenburg},\ and\
  \citenamefont {Smit}}]{venturoli2006mesoscopic}%
  \BibitemOpen
  \bibfield  {author} {\bibinfo {author} {\bibfnamefont {M.}~\bibnamefont
  {Venturoli}}, \bibinfo {author} {\bibfnamefont {M.~M.}\ \bibnamefont
  {Sperotto}}, \bibinfo {author} {\bibfnamefont {M.}~\bibnamefont
  {Kranenburg}}, \ and\ \bibinfo {author} {\bibfnamefont {B.}~\bibnamefont
  {Smit}},\ }\href@noop {} {\bibfield  {journal} {\bibinfo  {journal} {Phys.
  Rept.}\ }\textbf {\bibinfo {volume} {437}},\ \bibinfo {pages} {1} (\bibinfo
  {year} {2006})}\BibitemShut {NoStop}%
\bibitem [{\citenamefont {Watson}\ \emph {et~al.}(2013)\citenamefont {Watson},
  \citenamefont {Morriss-Andrews}, \citenamefont {Welch},\ and\ \citenamefont
  {Brown}}]{watson2013thermal}%
  \BibitemOpen
  \bibfield  {author} {\bibinfo {author} {\bibfnamefont {M.~C.}\ \bibnamefont
  {Watson}}, \bibinfo {author} {\bibfnamefont {A.}~\bibnamefont
  {Morriss-Andrews}}, \bibinfo {author} {\bibfnamefont {P.~M.}\ \bibnamefont
  {Welch}}, \ and\ \bibinfo {author} {\bibfnamefont {F.~L.}\ \bibnamefont
  {Brown}},\ }\href@noop {} {\bibfield  {journal} {\bibinfo  {journal} {J.
  Chem. Phys.}\ }\textbf {\bibinfo {volume} {139}},\ \bibinfo {pages} {084706}
  (\bibinfo {year} {2013})}\BibitemShut {NoStop}%
\bibitem [{\citenamefont {Argudo}\ \emph {et~al.}(2016)\citenamefont {Argudo},
  \citenamefont {Bethel}, \citenamefont {Marcoline},\ and\ \citenamefont
  {Grabe}}]{argudo2016continuum}%
  \BibitemOpen
  \bibfield  {author} {\bibinfo {author} {\bibfnamefont {D.}~\bibnamefont
  {Argudo}}, \bibinfo {author} {\bibfnamefont {N.~P.}\ \bibnamefont {Bethel}},
  \bibinfo {author} {\bibfnamefont {F.~V.}\ \bibnamefont {Marcoline}}, \ and\
  \bibinfo {author} {\bibfnamefont {M.}~\bibnamefont {Grabe}},\ }\href@noop {}
  {\bibfield  {journal} {\bibinfo  {journal} {Biochim. Biophys. Acta}\ }
  (\bibinfo {year} {2016})}\BibitemShut {NoStop}%
\bibitem [{\citenamefont {Nielsen}\ \emph {et~al.}(1998)\citenamefont
  {Nielsen}, \citenamefont {Goulian},\ and\ \citenamefont
  {Andersen}}]{nielsen1998energetics}%
  \BibitemOpen
  \bibfield  {author} {\bibinfo {author} {\bibfnamefont {C.}~\bibnamefont
  {Nielsen}}, \bibinfo {author} {\bibfnamefont {M.}~\bibnamefont {Goulian}}, \
  and\ \bibinfo {author} {\bibfnamefont {O.~S.}\ \bibnamefont {Andersen}},\
  }\href@noop {} {\bibfield  {journal} {\bibinfo  {journal} {Biophys. J.}\
  }\textbf {\bibinfo {volume} {74}},\ \bibinfo {pages} {1966} (\bibinfo {year}
  {1998})}\BibitemShut {NoStop}%
\bibitem [{\citenamefont {Ursell}\ \emph {et~al.}(2008)\citenamefont {Ursell},
  \citenamefont {Kondev}, \citenamefont {Reeves}, \citenamefont {Wiggins},\
  and\ \citenamefont {Phillips}}]{ursell2008role}%
  \BibitemOpen
  \bibfield  {author} {\bibinfo {author} {\bibfnamefont {T.}~\bibnamefont
  {Ursell}}, \bibinfo {author} {\bibfnamefont {J.}~\bibnamefont {Kondev}},
  \bibinfo {author} {\bibfnamefont {D.}~\bibnamefont {Reeves}}, \bibinfo
  {author} {\bibfnamefont {P.~A.}\ \bibnamefont {Wiggins}}, \ and\ \bibinfo
  {author} {\bibfnamefont {R.}~\bibnamefont {Phillips}},\ }in\ \href@noop {}
  {\emph {\bibinfo {booktitle} {Mechanosensitive Ion Channels}}}\ (\bibinfo
  {publisher} {Springer},\ \bibinfo {year} {2008})\ pp.\ \bibinfo {pages}
  {37--70}\BibitemShut {NoStop}%
\bibitem [{\citenamefont {Goulian}\ \emph {et~al.}(1993)\citenamefont
  {Goulian}, \citenamefont {Bruinsma},\ and\ \citenamefont
  {Pincus}}]{goulian1993long}%
  \BibitemOpen
  \bibfield  {author} {\bibinfo {author} {\bibfnamefont {M.}~\bibnamefont
  {Goulian}}, \bibinfo {author} {\bibfnamefont {R.}~\bibnamefont {Bruinsma}}, \
  and\ \bibinfo {author} {\bibfnamefont {P.}~\bibnamefont {Pincus}},\
  }\href@noop {} {\bibfield  {journal} {\bibinfo  {journal} {Europhys. Lett.}\
  }\textbf {\bibinfo {volume} {22}},\ \bibinfo {pages} {145} (\bibinfo {year}
  {1993})}\BibitemShut {NoStop}%
\bibitem [{\citenamefont {Netz}\ and\ \citenamefont
  {Pincus}(1995)}]{netz1995inhomogeneous}%
  \BibitemOpen
  \bibfield  {author} {\bibinfo {author} {\bibfnamefont {R.~R.}\ \bibnamefont
  {Netz}}\ and\ \bibinfo {author} {\bibfnamefont {P.}~\bibnamefont {Pincus}},\
  }\href@noop {} {\bibfield  {journal} {\bibinfo  {journal} {Phys. Rev. E}\
  }\textbf {\bibinfo {volume} {52}},\ \bibinfo {pages} {4114} (\bibinfo {year}
  {1995})}\BibitemShut {NoStop}%
\bibitem [{\citenamefont {Weikl}\ \emph {et~al.}(1998)\citenamefont {Weikl},
  \citenamefont {Kozlov},\ and\ \citenamefont
  {Helfrich}}]{weikl1998interaction}%
  \BibitemOpen
  \bibfield  {author} {\bibinfo {author} {\bibfnamefont {T.~R.}\ \bibnamefont
  {Weikl}}, \bibinfo {author} {\bibfnamefont {M.~M.}\ \bibnamefont {Kozlov}}, \
  and\ \bibinfo {author} {\bibfnamefont {W.}~\bibnamefont {Helfrich}},\
  }\href@noop {} {\bibfield  {journal} {\bibinfo  {journal} {Phys. Rev. E}\
  }\textbf {\bibinfo {volume} {57}},\ \bibinfo {pages} {6988} (\bibinfo {year}
  {1998})}\BibitemShut {NoStop}%
\bibitem [{\citenamefont {Chou}\ \emph {et~al.}(2001)\citenamefont {Chou},
  \citenamefont {Kim},\ and\ \citenamefont {Oster}}]{chou2001statistical}%
  \BibitemOpen
  \bibfield  {author} {\bibinfo {author} {\bibfnamefont {T.}~\bibnamefont
  {Chou}}, \bibinfo {author} {\bibfnamefont {K.~S.}\ \bibnamefont {Kim}}, \
  and\ \bibinfo {author} {\bibfnamefont {G.}~\bibnamefont {Oster}},\
  }\href@noop {} {\bibfield  {journal} {\bibinfo  {journal} {Biophys. J.}\
  }\textbf {\bibinfo {volume} {80}},\ \bibinfo {pages} {1075} (\bibinfo {year}
  {2001})}\BibitemShut {NoStop}%
\bibitem [{\citenamefont {Evans}\ \emph {et~al.}(2003)\citenamefont {Evans},
  \citenamefont {Turner},\ and\ \citenamefont {Sens}}]{evans2003interactions}%
  \BibitemOpen
  \bibfield  {author} {\bibinfo {author} {\bibfnamefont {A.~R.}\ \bibnamefont
  {Evans}}, \bibinfo {author} {\bibfnamefont {M.~S.}\ \bibnamefont {Turner}}, \
  and\ \bibinfo {author} {\bibfnamefont {P.}~\bibnamefont {Sens}},\ }\href@noop
  {} {\bibfield  {journal} {\bibinfo  {journal} {Phys. Rev. E}\ }\textbf
  {\bibinfo {volume} {67}},\ \bibinfo {pages} {041907} (\bibinfo {year}
  {2003})}\BibitemShut {NoStop}%
\bibitem [{\citenamefont {M{\"u}ller}\ \emph {et~al.}(2005)\citenamefont
  {M{\"u}ller}, \citenamefont {Deserno},\ and\ \citenamefont
  {Guven}}]{muller2005geometry}%
  \BibitemOpen
  \bibfield  {author} {\bibinfo {author} {\bibfnamefont {M.~M.}\ \bibnamefont
  {M{\"u}ller}}, \bibinfo {author} {\bibfnamefont {M.}~\bibnamefont {Deserno}},
  \ and\ \bibinfo {author} {\bibfnamefont {J.}~\bibnamefont {Guven}},\
  }\href@noop {} {\bibfield  {journal} {\bibinfo  {journal} {Europhys. Lett.}\
  }\textbf {\bibinfo {volume} {69}},\ \bibinfo {pages} {482} (\bibinfo {year}
  {2005})}\BibitemShut {NoStop}%
\bibitem [{\citenamefont {Haselwandter}\ and\ \citenamefont
  {Phillips}(2013)}]{haselwandter2013directional}%
  \BibitemOpen
  \bibfield  {author} {\bibinfo {author} {\bibfnamefont {C.~A.}\ \bibnamefont
  {Haselwandter}}\ and\ \bibinfo {author} {\bibfnamefont {R.}~\bibnamefont
  {Phillips}},\ }\href@noop {} {\bibfield  {journal} {\bibinfo  {journal}
  {Europhys. Lett.}\ }\textbf {\bibinfo {volume} {101}},\ \bibinfo {pages}
  {68002} (\bibinfo {year} {2013})}\BibitemShut {NoStop}%
\bibitem [{\citenamefont {Fournier}(2014)}]{fournier2014dynamics}%
  \BibitemOpen
  \bibfield  {author} {\bibinfo {author} {\bibfnamefont {J.-B.}\ \bibnamefont
  {Fournier}},\ }\href@noop {} {\bibfield  {journal} {\bibinfo  {journal}
  {Phys. Rev. Lett.}\ }\textbf {\bibinfo {volume} {112}},\ \bibinfo {pages}
  {128101} (\bibinfo {year} {2014})}\BibitemShut {NoStop}%
\bibitem [{\citenamefont {Reynwar}\ and\ \citenamefont
  {Deserno}(2011)}]{reynwar2011membrane}%
  \BibitemOpen
  \bibfield  {author} {\bibinfo {author} {\bibfnamefont {B.~J.}\ \bibnamefont
  {Reynwar}}\ and\ \bibinfo {author} {\bibfnamefont {M.}~\bibnamefont
  {Deserno}},\ }\href@noop {} {\bibfield  {journal} {\bibinfo  {journal} {Soft
  Matter}\ }\textbf {\bibinfo {volume} {7}},\ \bibinfo {pages} {8567} (\bibinfo
  {year} {2011})}\BibitemShut {NoStop}%
\bibitem [{\citenamefont {Paulsen}\ \emph {et~al.}(2015)\citenamefont
  {Paulsen}, \citenamefont {Armache}, \citenamefont {Gao}, \citenamefont
  {Cheng},\ and\ \citenamefont {Julius}}]{paulsen2015structure}%
  \BibitemOpen
  \bibfield  {author} {\bibinfo {author} {\bibfnamefont {C.~E.}\ \bibnamefont
  {Paulsen}}, \bibinfo {author} {\bibfnamefont {J.-P.}\ \bibnamefont
  {Armache}}, \bibinfo {author} {\bibfnamefont {Y.}~\bibnamefont {Gao}},
  \bibinfo {author} {\bibfnamefont {Y.}~\bibnamefont {Cheng}}, \ and\ \bibinfo
  {author} {\bibfnamefont {D.}~\bibnamefont {Julius}},\ }\href@noop {}
  {\bibfield  {journal} {\bibinfo  {journal} {Nature}\ }\textbf {\bibinfo
  {volume} {520}},\ \bibinfo {pages} {23} (\bibinfo {year} {2015})}\BibitemShut
  {NoStop}%
\bibitem [{\citenamefont {Cao}\ \emph {et~al.}(2013)\citenamefont {Cao},
  \citenamefont {Liao}, \citenamefont {Cheng},\ and\ \citenamefont
  {Julius}}]{cao2013trpv1}%
  \BibitemOpen
  \bibfield  {author} {\bibinfo {author} {\bibfnamefont {E.}~\bibnamefont
  {Cao}}, \bibinfo {author} {\bibfnamefont {M.}~\bibnamefont {Liao}}, \bibinfo
  {author} {\bibfnamefont {Y.}~\bibnamefont {Cheng}}, \ and\ \bibinfo {author}
  {\bibfnamefont {D.}~\bibnamefont {Julius}},\ }\href@noop {} {\bibfield
  {journal} {\bibinfo  {journal} {Nature}\ }\textbf {\bibinfo {volume} {504}},\
  \bibinfo {pages} {113} (\bibinfo {year} {2013})}\BibitemShut {NoStop}%
\bibitem [{\citenamefont {Brohawn}\ \emph {et~al.}(2014)\citenamefont
  {Brohawn}, \citenamefont {Su},\ and\ \citenamefont
  {MacKinnon}}]{brohawn2014mechanosensitivity}%
  \BibitemOpen
  \bibfield  {author} {\bibinfo {author} {\bibfnamefont {S.~G.}\ \bibnamefont
  {Brohawn}}, \bibinfo {author} {\bibfnamefont {Z.}~\bibnamefont {Su}}, \ and\
  \bibinfo {author} {\bibfnamefont {R.}~\bibnamefont {MacKinnon}},\ }\href@noop
  {} {\bibfield  {journal} {\bibinfo  {journal} {Proc. Nat. Acad. Sci.}\
  }\textbf {\bibinfo {volume} {111}},\ \bibinfo {pages} {3614} (\bibinfo {year}
  {2014})}\BibitemShut {NoStop}%
\bibitem [{\citenamefont {Lauritzen}\ \emph {et~al.}(2005)\citenamefont
  {Lauritzen}, \citenamefont {Chemin}, \citenamefont {Honore}, \citenamefont
  {Jodar}, \citenamefont {Guy}, \citenamefont {Lazdunski},\ and\ \citenamefont
  {Patel}}]{lauritzen2005cross}%
  \BibitemOpen
  \bibfield  {author} {\bibinfo {author} {\bibfnamefont {I.}~\bibnamefont
  {Lauritzen}}, \bibinfo {author} {\bibfnamefont {J.}~\bibnamefont {Chemin}},
  \bibinfo {author} {\bibfnamefont {E.}~\bibnamefont {Honore}}, \bibinfo
  {author} {\bibfnamefont {M.}~\bibnamefont {Jodar}}, \bibinfo {author}
  {\bibfnamefont {N.}~\bibnamefont {Guy}}, \bibinfo {author} {\bibfnamefont
  {M.}~\bibnamefont {Lazdunski}}, \ and\ \bibinfo {author} {\bibfnamefont
  {A.~J.}\ \bibnamefont {Patel}},\ }\href@noop {} {\bibfield  {journal}
  {\bibinfo  {journal} {EMBO rept.}\ }\textbf {\bibinfo {volume} {6}},\
  \bibinfo {pages} {642} (\bibinfo {year} {2005})}\BibitemShut {NoStop}%
\bibitem [{\citenamefont {Cox}\ \emph {et~al.}(2016)\citenamefont {Cox},
  \citenamefont {Bae}, \citenamefont {Ziegler}, \citenamefont {Hartley},
  \citenamefont {Nikolova-Krstevski}, \citenamefont {Rohde}, \citenamefont
  {Ng}, \citenamefont {Sachs}, \citenamefont {Gottlieb},\ and\ \citenamefont
  {Martinac}}]{cox2016removal}%
  \BibitemOpen
  \bibfield  {author} {\bibinfo {author} {\bibfnamefont {C.~D.}\ \bibnamefont
  {Cox}}, \bibinfo {author} {\bibfnamefont {C.}~\bibnamefont {Bae}}, \bibinfo
  {author} {\bibfnamefont {L.}~\bibnamefont {Ziegler}}, \bibinfo {author}
  {\bibfnamefont {S.}~\bibnamefont {Hartley}}, \bibinfo {author} {\bibfnamefont
  {V.}~\bibnamefont {Nikolova-Krstevski}}, \bibinfo {author} {\bibfnamefont
  {P.~R.}\ \bibnamefont {Rohde}}, \bibinfo {author} {\bibfnamefont {C.-A.}\
  \bibnamefont {Ng}}, \bibinfo {author} {\bibfnamefont {F.}~\bibnamefont
  {Sachs}}, \bibinfo {author} {\bibfnamefont {P.~A.}\ \bibnamefont {Gottlieb}},
  \ and\ \bibinfo {author} {\bibfnamefont {B.}~\bibnamefont {Martinac}},\
  }\href@noop {} {\bibfield  {journal} {\bibinfo  {journal} {Nature}\ }\textbf
  {\bibinfo {volume} {7}} (\bibinfo {year} {2016})}\BibitemShut {NoStop}%
\bibitem [{\citenamefont {Maksaev}\ \emph {et~al.}(2011)\citenamefont
  {Maksaev}, \citenamefont {Milac}, \citenamefont {Anishkin}, \citenamefont
  {Guy},\ and\ \citenamefont {Sukharev}}]{maksaev2011analyses}%
  \BibitemOpen
  \bibfield  {author} {\bibinfo {author} {\bibfnamefont {G.}~\bibnamefont
  {Maksaev}}, \bibinfo {author} {\bibfnamefont {A.}~\bibnamefont {Milac}},
  \bibinfo {author} {\bibfnamefont {A.}~\bibnamefont {Anishkin}}, \bibinfo
  {author} {\bibfnamefont {H.~R.}\ \bibnamefont {Guy}}, \ and\ \bibinfo
  {author} {\bibfnamefont {S.}~\bibnamefont {Sukharev}},\ }\href@noop {}
  {\bibfield  {journal} {\bibinfo  {journal} {Channels}\ }\textbf {\bibinfo
  {volume} {5}},\ \bibinfo {pages} {34} (\bibinfo {year} {2011})}\BibitemShut
  {NoStop}%
\end{thebibliography}

\newpage
\begin{thebibliography}{20}%
\makeatletter
\providecommand \@ifxundefined [1]{%
 \@ifx{#1\undefined}
}%
\providecommand \@ifnum [1]{%
 \ifnum #1\expandafter \@firstoftwo
 \else \expandafter \@secondoftwo
 \fi
}%
\providecommand \@ifx [1]{%
 \ifx #1\expandafter \@firstoftwo
 \else \expandafter \@secondoftwo
 \fi
}%
\providecommand \natexlab [1]{#1}%
\providecommand \enquote  [1]{``#1''}%
\providecommand \bibnamefont  [1]{#1}%
\providecommand \bibfnamefont [1]{#1}%
\providecommand \citenamefont [1]{#1}%
\providecommand \href@noop [0]{\@secondoftwo}%
\providecommand \href [0]{\begingroup \@sanitize@url \@href}%
\providecommand \@href[1]{\@@startlink{#1}\@@href}%
\providecommand \@@href[1]{\endgroup#1\@@endlink}%
\providecommand \@sanitize@url [0]{\catcode `\\12\catcode `\$12\catcode
  `\&12\catcode `\#12\catcode `\^12\catcode `\_12\catcode `\%12\relax}%
\providecommand \@@startlink[1]{}%
\providecommand \@@endlink[0]{}%
\providecommand \url  [0]{\begingroup\@sanitize@url \@url }%
\providecommand \@url [1]{\endgroup\@href {#1}{\urlprefix }}%
\providecommand \urlprefix  [0]{URL }%
\providecommand \Eprint [0]{\href }%
\providecommand \doibase [0]{http://dx.doi.org/}%
\providecommand \selectlanguage [0]{\@gobble}%
\providecommand \bibinfo  [0]{\@secondoftwo}%
\providecommand \bibfield  [0]{\@secondoftwo}%
\providecommand \translation [1]{[#1]}%
\providecommand \BibitemOpen [0]{}%
\providecommand \bibitemStop [0]{}%
\providecommand \bibitemNoStop [0]{.\EOS\space}%
\providecommand \EOS [0]{\spacefactor3000\relax}%
\providecommand \BibitemShut  [1]{\csname bibitem#1\endcsname}%
\let\auto@bib@innerbib\@empty
\bibitem [{\citenamefont {Ursell}\ \emph {et~al.}(2008)\citenamefont {Ursell},
  \citenamefont {Kondev}, \citenamefont {Reeves}, \citenamefont {Wiggins},\
  and\ \citenamefont {Phillips}}]{ursell2008role}%
  \BibitemOpen
  \bibfield  {author} {\bibinfo {author} {\bibfnamefont {T.}~\bibnamefont
  {Ursell}}, \bibinfo {author} {\bibfnamefont {J.}~\bibnamefont {Kondev}},
  \bibinfo {author} {\bibfnamefont {D.}~\bibnamefont {Reeves}}, \bibinfo
  {author} {\bibfnamefont {P.~A.}\ \bibnamefont {Wiggins}}, \ and\ \bibinfo
  {author} {\bibfnamefont {R.}~\bibnamefont {Phillips}},\ }in\ \href@noop {}
  {\emph {\bibinfo {booktitle} {Mechanosensitive Ion Channels}}}\ (\bibinfo
  {publisher} {Springer},\ \bibinfo {year} {2008})\ pp.\ \bibinfo {pages}
  {37--70}\BibitemShut {NoStop}%
\bibitem [{\citenamefont {Fournier}(2007)}]{fournier2007stress}%
  \BibitemOpen
  \bibfield  {author} {\bibinfo {author} {\bibfnamefont {J.-B.}\ \bibnamefont
  {Fournier}},\ }\href@noop {} {\bibfield  {journal} {\bibinfo  {journal} {Soft
  Matter}\ }\textbf {\bibinfo {volume} {3}},\ \bibinfo {pages} {883} (\bibinfo
  {year} {2007})}\BibitemShut {NoStop}%
\bibitem [{\citenamefont {Weikl}\ \emph {et~al.}(1998)\citenamefont {Weikl},
  \citenamefont {Kozlov},\ and\ \citenamefont
  {Helfrich}}]{weikl1998interaction}%
  \BibitemOpen
  \bibfield  {author} {\bibinfo {author} {\bibfnamefont {T.~R.}\ \bibnamefont
  {Weikl}}, \bibinfo {author} {\bibfnamefont {M.~M.}\ \bibnamefont {Kozlov}}, \
  and\ \bibinfo {author} {\bibfnamefont {W.}~\bibnamefont {Helfrich}},\
  }\href@noop {} {\bibfield  {journal} {\bibinfo  {journal} {Phys. Rev. E}\
  }\textbf {\bibinfo {volume} {57}},\ \bibinfo {pages} {6988} (\bibinfo {year}
  {1998})}\BibitemShut {NoStop}%
\bibitem [{\citenamefont {Nielsen}\ \emph {et~al.}(1998)\citenamefont
  {Nielsen}, \citenamefont {Goulian},\ and\ \citenamefont
  {Andersen}}]{nielsen1998energetics}%
  \BibitemOpen
  \bibfield  {author} {\bibinfo {author} {\bibfnamefont {C.}~\bibnamefont
  {Nielsen}}, \bibinfo {author} {\bibfnamefont {M.}~\bibnamefont {Goulian}}, \
  and\ \bibinfo {author} {\bibfnamefont {O.~S.}\ \bibnamefont {Andersen}},\
  }\href@noop {} {\bibfield  {journal} {\bibinfo  {journal} {Biophys. J.}\
  }\textbf {\bibinfo {volume} {74}},\ \bibinfo {pages} {1966} (\bibinfo {year}
  {1998})}\BibitemShut {NoStop}%
\bibitem [{\citenamefont {Huang}(1986)}]{huang1986deformation}%
  \BibitemOpen
  \bibfield  {author} {\bibinfo {author} {\bibfnamefont {H.~W.}\ \bibnamefont
  {Huang}},\ }\href@noop {} {\bibfield  {journal} {\bibinfo  {journal}
  {Biophys. J.}\ }\textbf {\bibinfo {volume} {50}},\ \bibinfo {pages} {1061}
  (\bibinfo {year} {1986})}\BibitemShut {NoStop}%
\bibitem [{\citenamefont {MacKintosh}\ and\ \citenamefont
  {Lubensky}(1991)}]{mackintosh1991orientational}%
  \BibitemOpen
  \bibfield  {author} {\bibinfo {author} {\bibfnamefont {F.}~\bibnamefont
  {MacKintosh}}\ and\ \bibinfo {author} {\bibfnamefont {T.}~\bibnamefont
  {Lubensky}},\ }\href@noop {} {\bibfield  {journal} {\bibinfo  {journal}
  {Phys. Rev. Lett.}\ }\textbf {\bibinfo {volume} {67}},\ \bibinfo {pages}
  {1169} (\bibinfo {year} {1991})}\BibitemShut {NoStop}%
\bibitem [{\citenamefont {Seifert}\ \emph {et~al.}(1996)\citenamefont
  {Seifert}, \citenamefont {Shillcock},\ and\ \citenamefont
  {Nelson}}]{seifert1996role}%
  \BibitemOpen
  \bibfield  {author} {\bibinfo {author} {\bibfnamefont {U.}~\bibnamefont
  {Seifert}}, \bibinfo {author} {\bibfnamefont {J.}~\bibnamefont {Shillcock}},
  \ and\ \bibinfo {author} {\bibfnamefont {P.}~\bibnamefont {Nelson}},\
  }\href@noop {} {\bibfield  {journal} {\bibinfo  {journal} {Phys. Rev. Lett.}\
  }\textbf {\bibinfo {volume} {77}},\ \bibinfo {pages} {5237} (\bibinfo {year}
  {1996})}\BibitemShut {NoStop}%
\bibitem [{\citenamefont {Fournier}(1999)}]{fournier1999microscopic}%
  \BibitemOpen
  \bibfield  {author} {\bibinfo {author} {\bibfnamefont {J.-B.}\ \bibnamefont
  {Fournier}},\ }\href@noop {} {\bibfield  {journal} {\bibinfo  {journal} {Eur.
  Phys. J. E}\ }\textbf {\bibinfo {volume} {11}},\ \bibinfo {pages} {261}
  (\bibinfo {year} {1999})}\BibitemShut {NoStop}%
\bibitem [{\citenamefont {Kozlovsky}\ \emph {et~al.}(2004)\citenamefont
  {Kozlovsky}, \citenamefont {Zimmerberg},\ and\ \citenamefont
  {Kozlov}}]{kozlovsky2004orientation}%
  \BibitemOpen
  \bibfield  {author} {\bibinfo {author} {\bibfnamefont {Y.}~\bibnamefont
  {Kozlovsky}}, \bibinfo {author} {\bibfnamefont {J.}~\bibnamefont
  {Zimmerberg}}, \ and\ \bibinfo {author} {\bibfnamefont {M.~M.}\ \bibnamefont
  {Kozlov}},\ }\href@noop {} {\bibfield  {journal} {\bibinfo  {journal}
  {Biophys. J.}\ }\textbf {\bibinfo {volume} {87}},\ \bibinfo {pages} {999}
  (\bibinfo {year} {2004})}\BibitemShut {NoStop}%
\bibitem [{\citenamefont {May}\ \emph {et~al.}(2004)\citenamefont {May},
  \citenamefont {Kozlovsky}, \citenamefont {Ben-Shaul},\ and\ \citenamefont
  {Kozlov}}]{may2004tilt}%
  \BibitemOpen
  \bibfield  {author} {\bibinfo {author} {\bibfnamefont {S.}~\bibnamefont
  {May}}, \bibinfo {author} {\bibfnamefont {Y.}~\bibnamefont {Kozlovsky}},
  \bibinfo {author} {\bibfnamefont {A.}~\bibnamefont {Ben-Shaul}}, \ and\
  \bibinfo {author} {\bibfnamefont {M.}~\bibnamefont {Kozlov}},\ }\href@noop {}
  {\bibfield  {journal} {\bibinfo  {journal} {Eur. Phys. J. E}\ }\textbf
  {\bibinfo {volume} {14}},\ \bibinfo {pages} {299} (\bibinfo {year}
  {2004})}\BibitemShut {NoStop}%
\bibitem [{\citenamefont {Argudo}\ \emph {et~al.}(2016)\citenamefont {Argudo},
  \citenamefont {Bethel}, \citenamefont {Marcoline},\ and\ \citenamefont
  {Grabe}}]{argudo2016continuum}%
  \BibitemOpen
  \bibfield  {author} {\bibinfo {author} {\bibfnamefont {D.}~\bibnamefont
  {Argudo}}, \bibinfo {author} {\bibfnamefont {N.~P.}\ \bibnamefont {Bethel}},
  \bibinfo {author} {\bibfnamefont {F.~V.}\ \bibnamefont {Marcoline}}, \ and\
  \bibinfo {author} {\bibfnamefont {M.}~\bibnamefont {Grabe}},\ }\href@noop {}
  {\bibfield  {journal} {\bibinfo  {journal} {Biochim. Biophys. Acta}\ }
  (\bibinfo {year} {2016})}\BibitemShut {NoStop}%
\bibitem [{\citenamefont {Jablin}\ \emph {et~al.}(2014)\citenamefont {Jablin},
  \citenamefont {Akabori},\ and\ \citenamefont
  {Nagle}}]{jablin2014experimental}%
  \BibitemOpen
  \bibfield  {author} {\bibinfo {author} {\bibfnamefont {M.}~\bibnamefont
  {Jablin}}, \bibinfo {author} {\bibfnamefont {K.}~\bibnamefont {Akabori}}, \
  and\ \bibinfo {author} {\bibfnamefont {J.}~\bibnamefont {Nagle}},\
  }\href@noop {} {\bibfield  {journal} {\bibinfo  {journal} {Phys. Rev. Lett.}\
  }\textbf {\bibinfo {volume} {113}},\ \bibinfo {pages} {248102} (\bibinfo
  {year} {2014})}\BibitemShut {NoStop}%
\bibitem [{\citenamefont {May}(2002)}]{may2002membrane}%
  \BibitemOpen
  \bibfield  {author} {\bibinfo {author} {\bibfnamefont {S.}~\bibnamefont
  {May}},\ }\href@noop {} {\bibfield  {journal} {\bibinfo  {journal}
  {Langmuir}\ }\textbf {\bibinfo {volume} {18}},\ \bibinfo {pages} {6356}
  (\bibinfo {year} {2002})}\BibitemShut {NoStop}%
\bibitem [{\citenamefont {Kuzmin}\ \emph {et~al.}(2005)\citenamefont {Kuzmin},
  \citenamefont {Akimov}, \citenamefont {Chizmadzhev}, \citenamefont
  {Zimmerberg},\ and\ \citenamefont {Cohen}}]{kuzmin2005line}%
  \BibitemOpen
  \bibfield  {author} {\bibinfo {author} {\bibfnamefont {P.~I.}\ \bibnamefont
  {Kuzmin}}, \bibinfo {author} {\bibfnamefont {S.~A.}\ \bibnamefont {Akimov}},
  \bibinfo {author} {\bibfnamefont {Y.~A.}\ \bibnamefont {Chizmadzhev}},
  \bibinfo {author} {\bibfnamefont {J.}~\bibnamefont {Zimmerberg}}, \ and\
  \bibinfo {author} {\bibfnamefont {F.~S.}\ \bibnamefont {Cohen}},\ }\href@noop
  {} {\bibfield  {journal} {\bibinfo  {journal} {Biophys. J.}\ }\textbf
  {\bibinfo {volume} {88}},\ \bibinfo {pages} {1120} (\bibinfo {year}
  {2005})}\BibitemShut {NoStop}%
\bibitem [{\citenamefont {Venturoli}\ \emph {et~al.}(2006)\citenamefont
  {Venturoli}, \citenamefont {Sperotto}, \citenamefont {Kranenburg},\ and\
  \citenamefont {Smit}}]{venturoli2006mesoscopic}%
  \BibitemOpen
  \bibfield  {author} {\bibinfo {author} {\bibfnamefont {M.}~\bibnamefont
  {Venturoli}}, \bibinfo {author} {\bibfnamefont {M.~M.}\ \bibnamefont
  {Sperotto}}, \bibinfo {author} {\bibfnamefont {M.}~\bibnamefont
  {Kranenburg}}, \ and\ \bibinfo {author} {\bibfnamefont {B.}~\bibnamefont
  {Smit}},\ }\href@noop {} {\bibfield  {journal} {\bibinfo  {journal} {Phys.
  Rept.}\ }\textbf {\bibinfo {volume} {437}},\ \bibinfo {pages} {1} (\bibinfo
  {year} {2006})}\BibitemShut {NoStop}%
\bibitem [{\citenamefont {Watson}\ \emph {et~al.}(2013)\citenamefont {Watson},
  \citenamefont {Morriss-Andrews}, \citenamefont {Welch},\ and\ \citenamefont
  {Brown}}]{watson2013thermal}%
  \BibitemOpen
  \bibfield  {author} {\bibinfo {author} {\bibfnamefont {M.~C.}\ \bibnamefont
  {Watson}}, \bibinfo {author} {\bibfnamefont {A.}~\bibnamefont
  {Morriss-Andrews}}, \bibinfo {author} {\bibfnamefont {P.~M.}\ \bibnamefont
  {Welch}}, \ and\ \bibinfo {author} {\bibfnamefont {F.~L.}\ \bibnamefont
  {Brown}},\ }\href@noop {} {\bibfield  {journal} {\bibinfo  {journal} {J.
  Chem. Phys.}\ }\textbf {\bibinfo {volume} {139}},\ \bibinfo {pages} {084706}
  (\bibinfo {year} {2013})}\BibitemShut {NoStop}%
\bibitem [{\citenamefont {Prager-Khoutorsky}\ \emph {et~al.}(2014)\citenamefont
  {Prager-Khoutorsky}, \citenamefont {Khoutorsky},\ and\ \citenamefont
  {Bourque}}]{prager2014unique}%
  \BibitemOpen
  \bibfield  {author} {\bibinfo {author} {\bibfnamefont {M.}~\bibnamefont
  {Prager-Khoutorsky}}, \bibinfo {author} {\bibfnamefont {A.}~\bibnamefont
  {Khoutorsky}}, \ and\ \bibinfo {author} {\bibfnamefont {C.~W.}\ \bibnamefont
  {Bourque}},\ }\href@noop {} {\bibfield  {journal} {\bibinfo  {journal}
  {Neuron}\ }\textbf {\bibinfo {volume} {83}},\ \bibinfo {pages} {866}
  (\bibinfo {year} {2014})}\BibitemShut {NoStop}%
\bibitem [{\citenamefont {Zhang}\ \emph {et~al.}(2015)\citenamefont {Zhang},
  \citenamefont {Cheng}, \citenamefont {Kittelmann}, \citenamefont {Li},
  \citenamefont {Petkovic}, \citenamefont {Cheng}, \citenamefont {Jin},
  \citenamefont {Guo}, \citenamefont {G{\"o}pfert}, \citenamefont {Jan} \emph
  {et~al.}}]{zhang2015ankyrin}%
  \BibitemOpen
  \bibfield  {author} {\bibinfo {author} {\bibfnamefont {W.}~\bibnamefont
  {Zhang}}, \bibinfo {author} {\bibfnamefont {L.~E.}\ \bibnamefont {Cheng}},
  \bibinfo {author} {\bibfnamefont {M.}~\bibnamefont {Kittelmann}}, \bibinfo
  {author} {\bibfnamefont {J.}~\bibnamefont {Li}}, \bibinfo {author}
  {\bibfnamefont {M.}~\bibnamefont {Petkovic}}, \bibinfo {author}
  {\bibfnamefont {T.}~\bibnamefont {Cheng}}, \bibinfo {author} {\bibfnamefont
  {P.}~\bibnamefont {Jin}}, \bibinfo {author} {\bibfnamefont {Z.}~\bibnamefont
  {Guo}}, \bibinfo {author} {\bibfnamefont {M.~C.}\ \bibnamefont
  {G{\"o}pfert}}, \bibinfo {author} {\bibfnamefont {L.~Y.}\ \bibnamefont
  {Jan}},  \emph {et~al.},\ }\href@noop {} {\bibfield  {journal} {\bibinfo
  {journal} {Cell}\ }\textbf {\bibinfo {volume} {162}},\ \bibinfo {pages}
  {1391} (\bibinfo {year} {2015})}\BibitemShut {NoStop}%
\bibitem [{\citenamefont {Hayakawa}\ \emph {et~al.}(2008)\citenamefont
  {Hayakawa}, \citenamefont {Tatsumi},\ and\ \citenamefont
  {Sokabe}}]{hayakawa2008actin}%
  \BibitemOpen
  \bibfield  {author} {\bibinfo {author} {\bibfnamefont {K.}~\bibnamefont
  {Hayakawa}}, \bibinfo {author} {\bibfnamefont {H.}~\bibnamefont {Tatsumi}}, \
  and\ \bibinfo {author} {\bibfnamefont {M.}~\bibnamefont {Sokabe}},\
  }\href@noop {} {\bibfield  {journal} {\bibinfo  {journal} {J. Cell Sci.}\
  }\textbf {\bibinfo {volume} {121}},\ \bibinfo {pages} {496} (\bibinfo {year}
  {2008})}\BibitemShut {NoStop}%
\bibitem [{\citenamefont {Maksaev}\ \emph {et~al.}(2011)\citenamefont
  {Maksaev}, \citenamefont {Milac}, \citenamefont {Anishkin}, \citenamefont
  {Guy},\ and\ \citenamefont {Sukharev}}]{maksaev2011analyses}%
  \BibitemOpen
  \bibfield  {author} {\bibinfo {author} {\bibfnamefont {G.}~\bibnamefont
  {Maksaev}}, \bibinfo {author} {\bibfnamefont {A.}~\bibnamefont {Milac}},
  \bibinfo {author} {\bibfnamefont {A.}~\bibnamefont {Anishkin}}, \bibinfo
  {author} {\bibfnamefont {H.~R.}\ \bibnamefont {Guy}}, \ and\ \bibinfo
  {author} {\bibfnamefont {S.}~\bibnamefont {Sukharev}},\ }\href@noop {}
  {\bibfield  {journal} {\bibinfo  {journal} {Channels}\ }\textbf {\bibinfo
  {volume} {5}},\ \bibinfo {pages} {34} (\bibinfo {year} {2011})}\BibitemShut
  {NoStop}%
\end{thebibliography}
%
\end{widetext}
\end{document}